\documentclass[twocolumn,showpacs,showkeys,preprintnumbers,amsmath,amssymb,floatfix,aps,prc,10pt,colorlinks,urlcolor=blue]{revtex4-1}
\usepackage[latin9]{inputenc}

\usepackage{feynmp-auto}
\usepackage{color}
\usepackage{graphicx}
\usepackage{float}
\usepackage{mathtools}
\usepackage{makecell}
\usepackage{multirow}
\usepackage{siunitx}
\usepackage{array,booktabs,tabularx,longtable}
\usepackage{afterpage} 
\usepackage[colorlinks,urlcolor=blue]{hyperref}
\usepackage{breakurl}

\usepackage[normalem]{ulem}
\newcounter{univ_counter}
\addtocounter{univ_counter} {1} 
\edef\ANL{$^{\arabic{univ_counter}}$ }

\addtocounter{univ_counter} {1} 
\edef\ASU{$^{\arabic{univ_counter}}$ }

\addtocounter{univ_counter} {1} 
\edef\BRESCIA{$^{\arabic{univ_counter}}$ }

\addtocounter{univ_counter} {1} 
\edef\UCR{$^{\arabic{univ_counter}}$ }

\addtocounter{univ_counter} {1} 
\edef\CSUDH{$^{\arabic{univ_counter}}$ }

\addtocounter{univ_counter} {1} 
\edef\CANISIUS{$^{\arabic{univ_counter}}$ }

\addtocounter{univ_counter} {1} 
\edef\CUA{$^{\arabic{univ_counter}}$ }

\addtocounter{univ_counter} {1} 
\edef\CMU{$^{\arabic{univ_counter}}$ }

\addtocounter{univ_counter} {1} 
\edef\CNU{$^{\arabic{univ_counter}}$ }

\addtocounter{univ_counter} {1} 
\edef\UCONN{$^{\arabic{univ_counter}}$ }

\addtocounter{univ_counter} {1} 
\edef\DUKE{$^{\arabic{univ_counter}}$ }

\addtocounter{univ_counter} {1} 
\edef\DUQUESNE{$^{\arabic{univ_counter}}$ }

\addtocounter{univ_counter} {1} 
\edef\FU{$^{\arabic{univ_counter}}$ }

\addtocounter{univ_counter} {1}
\edef\FERRARAU{$^{\arabic{univ_counter}}$ }

\addtocounter{univ_counter} {1} 
\edef\FIU{$^{\arabic{univ_counter}}$ }

\addtocounter{univ_counter} {1} 
\edef\FSU{$^{\arabic{univ_counter}}$ }

\addtocounter{univ_counter} {1}
\edef\GSIFFN{$^{\arabic{univ_counter}}$ }

\addtocounter{univ_counter} {1} 
\edef\GWUI{$^{\arabic{univ_counter}}$ }

\addtocounter{univ_counter} {1} 
\edef\GLASGOW{$^{\arabic{univ_counter}}$ }

\addtocounter{univ_counter} {1} 
\edef\INFNCAT{$^{\arabic{univ_counter}}$ }

\addtocounter{univ_counter} {1} 
\edef\INFNFE{$^{\arabic{univ_counter}}$ }

\addtocounter{univ_counter} {1} 
\edef\INFNFR{$^{\arabic{univ_counter}}$ }

\addtocounter{univ_counter} {1} 
\edef\INFNGE{$^{\arabic{univ_counter}}$ }

\addtocounter{univ_counter} {1} 
\edef\INFNPAV{$^{\arabic{univ_counter}}$ }

\addtocounter{univ_counter} {1} 
\edef\INFNRO{$^{\arabic{univ_counter}}$ }

\addtocounter{univ_counter} {1} 
\edef\INFNTUR{$^{\arabic{univ_counter}}$ }

\addtocounter{univ_counter} {1} 
\edef\ISU{$^{\arabic{univ_counter}}$ }

\addtocounter{univ_counter} {1} 
\edef\JLUGiessen{$^{\arabic{univ_counter}}$ }

\addtocounter{univ_counter} {1} 
\edef\JMU{$^{\arabic{univ_counter}}$ }

\addtocounter{univ_counter} {1} 
\edef\KSU{$^{\arabic{univ_counter}}$ }

\addtocounter{univ_counter} {1} 
\edef\KNU{$^{\arabic{univ_counter}}$ }

\addtocounter{univ_counter} {1} 
\edef\LAMAR{$^{\arabic{univ_counter}}$ }

\addtocounter{univ_counter} {1} 
\edef\MIT{$^{\arabic{univ_counter}}$ }

\addtocounter{univ_counter} {1} 
\edef\MISS{$^{\arabic{univ_counter}}$ }

\addtocounter{univ_counter} {1} 
\edef\UNH{$^{\arabic{univ_counter}}$ }

\addtocounter{univ_counter} {1} 
\edef\NMSU{$^{\arabic{univ_counter}}$ }

\addtocounter{univ_counter} {1} 
\edef\NSU{$^{\arabic{univ_counter}}$ }

\addtocounter{univ_counter} {1} 
\edef\OHIOU{$^{\arabic{univ_counter}}$ }

\addtocounter{univ_counter} {1} 
\edef\ODU{$^{\arabic{univ_counter}}$ }

\addtocounter{univ_counter} {1} 
\edef\ORSAY{$^{\arabic{univ_counter}}$ }

\addtocounter{univ_counter} {1} 
\edef\URICH{$^{\arabic{univ_counter}}$ }

\addtocounter{univ_counter} {1} 
\edef\ROMAII{$^{\arabic{univ_counter}}$ }

\addtocounter{univ_counter} {1} 
\edef\RPI{$^{\arabic{univ_counter}}$ }

\addtocounter{univ_counter} {1} 
\edef\SACLAY{$^{\arabic{univ_counter}}$ }

\addtocounter{univ_counter} {1} 
\edef\MSU{$^{\arabic{univ_counter}}$ }

\addtocounter{univ_counter} {1} 
\edef\SCAROLINA{$^{\arabic{univ_counter}}$ }

\addtocounter{univ_counter} {1} 
\edef\TEMPLE{$^{\arabic{univ_counter}}$ }

\addtocounter{univ_counter} {1} 
\edef\UTK{$^{\arabic{univ_counter}}$ }

\addtocounter{univ_counter} {1} 
\edef\UTFSM{$^{\arabic{univ_counter}}$ }

\addtocounter{univ_counter} {1}
\edef\JLAB{$^{\arabic{univ_counter}}$ }

\addtocounter{univ_counter} {1}
\edef\ULS{$^{\arabic{univ_counter}}$ }

\addtocounter{univ_counter} {1} 
\edef\VIRGINIA{$^{\arabic{univ_counter}}$ }

\addtocounter{univ_counter} {1} 
\edef\VT{$^{\arabic{univ_counter}}$ }

\addtocounter{univ_counter} {1} 
\edef\WM{$^{\arabic{univ_counter}}$ }

\addtocounter{univ_counter} {1} 
\edef\YEREVAN{$^{\arabic{univ_counter}}$ }

\addtocounter{univ_counter} {1} 
\edef\YORK{$^{\arabic{univ_counter}}$ }

\begin{document}

\preprint{Phys. Rev. C}

\title{Measurement of Beam-Recoil Observables $C_x$ and $C_z$ for $K^+\Lambda$ Photoproduction}


\author{
S. Adhikari,\FIU\
B.A.~Raue$^\dag$,\FIU\
D.S.~Carman$^\dag$,\JLAB\
L.~Guo,\FIU\
T. Chetry,\FIU\
P.~Achenbach,\JLAB\
J.S.~Alvarado,\ORSAY\
M.J.~Amaryan,\ODU\
W.R.~Armstrong,\ANL\
H.~Atac,\TEMPLE
% ...
\\
(CLAS Collaboration)}

\affiliation{\ANL Argonne National Laboratory, Argonne, Illinois 60439}
\affiliation{\ASU ASU, Arizona State University, Tempe, Arizona 85287}
\affiliation{\BRESCIA Universit\`{a} degli Studi di Brescia, 25123 Brescia, Italy}
\affiliation{\UCR University of California Riverside, Riverside, California 92521}
\affiliation{\CSUDH California State University, Dominguez Hills, Carson, California 90747}
\affiliation{\CANISIUS Canisius College, Buffalo, New York 14208}
\affiliation{\CUA Catholic University of America, Washington, D.C. 20064}
\affiliation{\CMU Carnegie Mellon University, Pittsburgh, Pennsylvania 15213}
\affiliation{\CNU Christopher Newport University, Newport News, Virginia 23606}
\affiliation{\UCONN University of Connecticut, Storrs, Connecticut 06269}
\affiliation{\DUKE Duke University, Durham, North Carolina 27708}
\affiliation{\DUQUESNE Duquesne University, Pittsburgh, Pennsylvania 15282}
\affiliation{\FU Fairfield University, Fairfield, Connecticut 06824}
\affiliation{\FERRARAU Universit\`{a} di Ferrara, 44121 Ferrara, Italy}
\affiliation{\FIU Florida International University, Miami, Florida 33199}
\affiliation{\FSU Florida State University, Tallahassee, Florida 32306}
\affiliation{\GSIFFN GSI Helmholtzzentrum fur Schwerionenforschung GmbH, D-64291 Darmstadt, Germany}
\affiliation{\GWUI The George Washington University, Washington, D.C. 20052}
\affiliation{\GLASGOW University of Glasgow, Glasgow G12 8QQ, United Kingdom}
\affiliation{\INFNCAT INFN, Sezione di Catania, 95123 Catania, Italy}
\affiliation{\INFNFE INFN, Sezione di Ferrara, 44100 Ferrara, Italy}
\affiliation{\INFNFR INFN, Laboratori Nazionali di Frascati, 00044 Frascati, Italy}
\affiliation{\INFNGE INFN, Sezione di Genova, 16146 Genova, Italy}
\affiliation{\INFNPAV INFN, Sezione di Pavia, 27100 Pavia, Italy}
\affiliation{\INFNRO INFN, Sezione di Roma Tor Vergata, 00133 Rome, Italy}
\affiliation{\INFNTUR INFN, Sezione di Torino, 10125 Torino, Italy}
\affiliation{\ISU Idaho State University, Pocatello, Idaho 83209}
\affiliation{\JLUGiessen II Physikalisches Institut der Universitaet Giessen, 35392 Giessen, Germany}
\affiliation{\JMU James Madison University, Harrisonburg, Virginia 22807}
\affiliation{\KSU King Saud University, Riyadh, 11362 Kingdom of Saudi Arabia}
\affiliation{\KNU Kyungpook National University, Daegu 702-701, Republic of Korea}
\affiliation{\LAMAR Lamar University, Beaumont, Texas 77710}
\affiliation{\MIT Massachusetts Institute of Technology, Cambridge, Massachusetts 02139}
\affiliation{\MISS Mississippi State University, Mississippi State, Mississippi 39762}
\affiliation{\UNH University of New Hampshire, Durham, New Hampshire 03824}
\affiliation{\NMSU New Mexico State University, Las Cruces, New Mexico 88003}
\affiliation{\NSU Norfolk State University, Norfolk, Virginia 23504}
\affiliation{\OHIOU Ohio University, Athens, Ohio 45701}
\affiliation{\ODU Old Dominion University, Norfolk, Virginia 23529}
\affiliation{\ORSAY Universit\'{e} Paris-Saclay, CNRS/IN2P3, IJCLab, 91405 Orsay, France}
\affiliation{\URICH University of Richmond, Richmond, Virginia 23173}
\affiliation{\ROMAII Universit\`{a} di Roma Tor Vergata, 00133 Rome, Italy}
\affiliation{\RPI Rensselaer Polytechnic Institute, Troy, New York 12180}
\affiliation{\SACLAY IRFU, CEA, Universit\'{e} Paris-Saclay, F-91191 Gif-sur-Yvette, France}
\affiliation{\MSU Skobeltsyn Nuclear Physics Institute and Physics Department at Lomonosov Moscow State University, 119899 Moscow, Russia}
\affiliation{\SCAROLINA University of South Carolina, Columbia, South Carolina 29208}
\affiliation{\TEMPLE Temple University, Philadelphia, Pennsylvania 19122}
\affiliation{\UTK University of Tennessee, Knoxville, Tennessee 37996}
\affiliation{\UTFSM Universidad T\'{e}cnica Federico Santa Mar\'{i}a, Casilla 110-V Valpara\'{i}so, Chile}
\affiliation{\JLAB Thomas Jefferson National Accelerator Facility, Newport News, Virginia 23606}
\affiliation{\ULS Universidad de La Serena, Avda. Juan Cisternas 1200, La Serena, Chile}
\affiliation{\VIRGINIA University of Virginia, Charlottesville, Virginia 22901}
\affiliation{\VT Virginia Tech, Blacksburg, Virginia 24061}
\affiliation{\WM College of William and Mary, Williamsburg, Virginia 23187}
\affiliation{\YEREVAN Yerevan Physics Institute, 375036 Yerevan, Armenia}
\affiliation{\YORK University of York, York YO10 5DD, United Kingdom}

\date{\today}

\begin{abstract}
Exclusive photoproduction of $K^+ \Lambda$ final states off a proton target has been an important component in the search for missing nucleon resonances 
and our understanding of the production of final states containing strange quarks. Polarization observables have been instrumental in this effort. The current 
work is an extension of previously published CLAS results on the beam-recoil transferred polarization observables $C_x$ and $C_z$. We extend the kinematic range 
up to invariant mass $W=3.33$~GeV from the previous limit of $W=2.5$~GeV with significantly improved statistical precision in the region of overlap. These data 
will provide for tighter constraints on the reaction models used to unravel the spectrum of nucleon resonances and their properties by not only improving the statistical 
precision of the data within the resonance region, but also by constraining $t$-channel processes that dominate at higher $W$ but extend into the resonance region.
\end{abstract}

\maketitle

\section{Introduction}
\label{sec:intro}

During the past 20 years a number of groups have collected high-quality data that have provided for a broad array of observables from exclusive meson photo- 
and electroproduction experiments. This advancement has allowed for marked progress in mapping out the spectrum of the excited states of the nucleon ($N^*$s) 
and in understanding their internal structure. It is  information from these resonance states that encodes the underlying dynamics of these strongly 
coupled systems and allows for detailed insights into Quantum Chromodynamics (QCD) in the non-perturbative domain. The majority of the forward progress in 
this area has been provided by advanced analyses of the $\pi N$ and $\pi^+ \pi^- p$ exclusive reaction channels~\cite{dsc-fbs2020,symmetry2025}. 

However, in recent years the high-precision data from CLAS on exclusive photoproduction of $K^+Y$ ($Y=\Lambda,\Sigma^0$) 
\cite{mcnabb,bradford2006,bradford2007,mccracken2010,dey2010,paterson2016} have proven crucial in this advancement, in particular, for the discovery of new baryon 
states known as the ``missing" resonances. In the strangeness photoproduction channels, data are also available from MAMI \cite{mami1}, SAPHIR~\cite{saphir1}, 
GRAAL~\cite{graal1,graal2}, LEPS~\cite{leps1,leps2}, and BGO-OD~\cite{bgood1}. $KY$ exclusive production is sensitive to coupling to higher-lying $N^*$ states for 
invariant mass $W > 1.6$~GeV, which is precisely the mass range where the understanding of the $N^*$ spectrum is most limited. With these data, roughly a dozen 
$N^*$ states have been confirmed within global multi-channel analyses of the exclusive photoproduction data with a decisive impact from the hyperon 
polarization observables~\cite{Bur17,Burkert:2020}. 

Gaining insight into QCD in the non-perturbative regime requires not only high-precision experimental data, but advanced reaction models that accurately describe
the data over a broad kinematic range. These reaction models include single-channel models \cite{fbs-mokeev,kaon-maid,rpr-bs,maxwell,skoupil18} and dynamical 
coupled-channel models~\cite{anl-osaka,jb-model,bnga2012}. However, the constraints and insights that such phenomenological reaction models provide to understand strong 
QCD dynamics are only as good as the quality of the experimental data. The less precise and complete the available data, the larger the uncertainties and ambiguities 
in the extracted multipoles, photocouplings, and decay widths that encode the nucleon resonance structure information. Improving the statistical and systematic precision, 
as well as extending the kinematic range of the available data will be critical to foster these efforts and to further understand both the resonant and non-resonant 
contributions to the reaction mechanism.

The current work is an extension of previously published CLAS photoproduction data on the $\Lambda$ hyperon polarization observables $C_x$ and $C_z$ \cite{bradford2007}. 
These observables characterize the polarization transfer from a circularly polarized incident photon beam to a recoiling hyperon along orthogonal axes in the 
hadron reaction plane. We extend the kinematic range up to invariant mass $W=3.33$~GeV from the previous limit of $W=2.5$~GeV with significantly improved 
statistical precision in the region of overlap ($1.78 < W < 2.5$~GeV). While the present results extend beyond what is generally considered the resonance 
region, it is important to understand this higher $W$ region as it is expected to be dominated by non-resonant $t$-channel processes. Such processes extend into the 
resonance region and are considered to be an underlying background. Therefore, understanding the $t$-channel contributions at higher $W$ and extrapolating them into the 
resonance region provides an important constraint on missing resonance identification.

The organization for the remainder of this paper is as follows. In Section~\ref{sec:formalism} the formalism of the beam-recoil transferred polarization and the
polarization extraction approach are presented. Section~\ref{sec:details} provides details on the experiment, describes the event selection procedures, data 
binning, yield extraction method, and discussion of the systematic uncertainty analysis. Section~\ref{sec:results} presents the measured $\Lambda$ hyperon $C_x$ and 
$C_z$ observables compared with available data and several model predictions. Finally, a summary of this work and our conclusions are given in 
Section~\ref{sec:conclusions}.

\section{Formalism and Extraction Method}
\label{sec:formalism}

\subsection{Observable Definitions}
\label{sec-observables}

The $K^+Y$ cross section for producing a polarized hyperon $\vec Y$ from a circularly polarized photon beam on an unpolarized proton target is given by 
\cite{Barker:1975bp}
\begin{equation}
\label{eq-Pcsec}
\rho_Y\frac{d\sigma}{d\Omega_K} = \frac{d\sigma}{d\Omega_K} \Big|_{unpol}\bigl(1 + \sigma_y P + P_{\odot}\left(C_x\sigma_x + C_z\sigma_z\right)\bigr),
\end{equation}
\noindent
where the hyperon density matrix $\rho_{Y}$ is
\begin{equation}
\label{Eq:densityMatrix}
\rho_Y = \left( 1 + \vec{\sigma}\cdot\vec{P_{Y}}\right),
\end{equation}
\noindent
with $\vec{P}_Y$ the hyperon polarization and $\vec{\sigma}$ the Pauli spin matrices.

The observables in Eq.(\ref{eq-Pcsec}) are the recoil polarization $P$ (not extracted in the present analysis) and the polarization transfer coefficients
$C_x$ and $C_z$. $P_{\odot}$ is the energy-dependent circular polarization of the photons originating from the bremsstrahlung of the longitudinally polarized
electrons on a radiator.

\begin{figure}[tb]
\centering
\includegraphics[width=0.9\columnwidth]{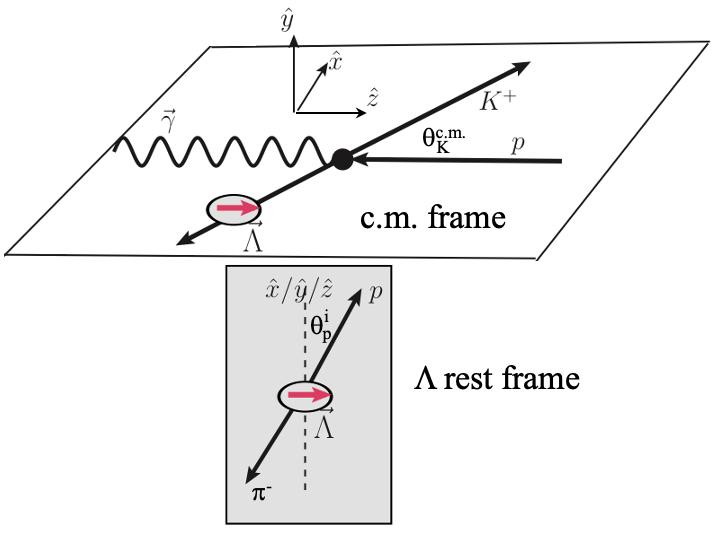}
\caption{Center-of-mass coordinate system definition for measurement of the observables $P$, $C_x$, and $C_z$ for the $\Lambda$. The recoiling $\Lambda$ is 
written as a vector to represent polarized production. The shaded box represents the $\Lambda$ rest frame showing the rest-frame decay angle $\theta_p^i$ 
of the proton with respect to the chosen axis $i = x, y, z$.}
\label{fig:axisDef}
\end{figure}

The recoil polarization $P$ can be measured by determining the $y$ component of the $\Lambda$ polarization, where $\hat y$ is the normal to the production 
plane in the beam photon + target proton center-of-mass (c.m.) frame. In this system the $z$-axis is along the direction of the beam photon and the $x$-axis
forms a right-handed coordinate system (see Fig.~\ref{fig:axisDef}). The products of the transferred polarizations $C_x$ and $C_z$ and the photon beam polarization 
$P_\odot$ are thus the projections of the $\Lambda$ polarization along the $x$ and $z$ axes, respectively, expressed as

\begin{eqnarray}
\label{eq-AllP}
P_{Y_x} &=& P_\odot C_x \\ \nonumber
P_{Y_y} &=& P \\ \nonumber
P_{Y_z} &=& P_\odot C_z.
\end{eqnarray}

The axis convention chosen in this work is consistent with that of Ref.~\cite{bradford2007}, where $C_x$ and $C_z$ are defined with opposite signs compared to 
Eq.(\ref{eq-Pcsec}). This ensures that $C_z$ is positive when the two $z$ axes in the c.m.~frame and $\Lambda$ rest frame coincide at the forward $K^+$ production 
angle. In other words, the hyperon polarization is positive along the $z$-axis.

\subsection{Observable Extraction Using Maximum Likelihood Method}
\label{sec-MLM}

The hyperon polarization is related to the decay proton angular distributions in the hyperon rest frame given by

\begin{equation}
\label{eq-angdist}
I_i(\cos\theta_p^i) = \frac{1}{2}\left(1+\alpha P_{Y_i}\cos\theta_p^i\right),
\end{equation}

\noindent
where $\theta_p^i$ is the proton angle with respect to a given coordinate axis, $i\in{x,y,z}$. The weak decay asymmetry parameter is $\alpha=0.747\pm 0.007$
\cite{pdg2024}. 

In principle, it is possible to do a simultaneous extraction of $P$, $C_x$, and $C_z$ using the maximum likelihood method (MLM)~\cite{IrelandMLM}. The likelihood 
function is a product of probability density functions ({\tt pdf})
\begin{equation}
\label{eq-lf}
{\cal L}=\prod_{i=1}^N {\cal P}_i,
\end{equation}
\noindent
where the {\tt pdf} for the case of simultaneous extraction of $P$, $C_x$, and $C_z$ is
\begin{multline}
\label{eq-pdf}
{\cal P}(\cos\theta_p^x,\cos\theta_p^y,\cos\theta_p^z|C_x,C_z,P) = 1\pm \alpha P_\odot \\ 
\left(C_x \cos\theta_p^x+C_z\cos\theta_p^z\right) + \alpha P\cos\theta_p^y.
\end{multline}
\noindent
Here, the cosine terms are the variables in the {\tt pdf}, and $C_x$, $C_z$, and $P$ are the parameters to be determined. For a given kinematic bin there are $N$ 
events, each with weight $w_i$ and helicity of $+$ or $-$. One then minimizes the negative log-likelihood given by
\begin{multline}
\label{eq-logL}
-\log{\cal L} = -\sum_i^N w_i \log\bigl(P_\odot\alpha \left(C_x \cos\theta_p^x+C_z\cos\theta_p^z\right) \\ + ~\alpha P\cos\theta_p^y\bigr).
\end{multline}
\noindent
This formulation requires the weight to be defined as the product of the event's $Q$-factor and the acceptance correction factor: $w_i=Q_i f_{acc}^i$ (see
Section~\ref{sec-Q} for details on the definition of the $Q$ factor). Since the decay-proton acceptance is not uniform over the $\cos \theta_p^y$ range, then 
corrections need to be applied to extract $P$. However, to leading order, acceptance corrections are not necessary for the extraction of $C_x$ and $C_z$ since 
the acceptance for positive and negative helicity events is equivalent (apart from finite-resolution effects). 

In this work, only $C_x$ and $C_z$ are extracted using $w_i=Q_i$ and a {\tt pdf} given by
\begin{multline}
\label{eq-Cpdf}
{\cal P}_C (\cos\theta_p^x,\cos\theta_p^z|C_x,C_z) = \\
1 \pm \alpha P_\odot \left(C_x \cos\theta_p^x + C_z \cos\theta_p^z \right).
\end{multline}
\noindent
This leads to
\begin{equation}
\label{eq-logLC}
-\log{\cal L}_C=-\sum_i^N Q_i \log\bigl(P_\odot\alpha\left(C_x \cos\theta_p^x+C_z\cos\theta_p^z\right)\bigr).
\end{equation}

In this fitting approach there are no constraints built-in to enforce $\vert C_x \vert, \vert C_z \vert \le 1$. Thus extractions of the hyperon polarization
outside of the physical bounds is possible due to statistical and/or systematic effects.

\subsection{Alternative Extraction Procedure}

Some prior extractions of hyperon polarization \cite{bradford2007,BONO2018280} have used a method of binning the beam-helicity asymmetries
in $\cos \theta_p$ and doing a 1-dimensional linear fit to

\begin{equation}
\label{eq-1Dfit}
A(\cos\theta_p^{x/z}) = \frac{N_{+} - N_{-}}{N_{+} - N_{-}} = \alpha P_\odot C_{x/z}\cos\theta_p^{x/z}.
\end{equation}

\noindent
This introduces potential systematic uncertainties due to the choice in binning (see Ref.~\cite{carman2003}). Since the MLM fit is done event-by-event (unbinned), 
it does not have this systematic uncertainty. This is also why it is not possible to show a fit for the MLM. However, comparisons of results from the 
1D linear fits and 2D MLM fits show they are fully consistent over the vast majority of the bins (see Section~\ref{sec:results}).

\section{Experimental Details}
\label{sec:details}

Electrons from the CEBAF accelerator at Jefferson Laboratory with a beam energy of 5.715~GeV were directed onto a thin gold radiator foil to produce bremsstrahlung
photons that were collimated onto a 40-cm-long liquid-hydrogen target. The CLAS detector~\cite{clas-nim} was used for this experiment, which was part of the
g12 Run Group that took data in 2008. The target center was 90-cm upstream from the nominal center of the CLAS detector to provide better acceptance for 
charged particles produced at small angles. To allow for high luminosity with a beam current of 60-65~nA, a 24-segment scintillator start counter~\cite{st-counter}
(ST) around the target was used to form a coincidence trigger with the scintillator counters of the time-of-flight (TOF) system~\cite{clas-tof} that surrounded the 
outside of CLAS. A coincidence of two ST-TOF hits in separate sectors of the CLAS detector in conjunction with a scattered electron detected in the bremsstrahlung
tagger~\cite{clas-tagger} was required to satisfy the trigger. These conditions, along with several ancillary trigger definitions, resulted in a live time of 
the data-acquisition system of $\sim$87\%.

The Hall~B M{\o}ller polarimeter is a dual-arm coincidence device that exploits the helicity dependence of $e^- + e^-$ scattering to measure the polarization of 
the incident electron beam. As the operation of the polarimeter disrupted the beam, beam polarization measurements were performed periodically within the main 
data-taking sequence throughout the run period. The average electron beam polarization for the g12 run period was 70\% with a relative uncertainty estimated to be 5\%. 
The polarization of the beam was flipped at the injector to the accelerator at a rate of 30~Hz in a simple non-random $+ - + -$ sequence. The beam helicity 
state was recorded event-by-event in the data stream. The energy-dependent circular polarization $P_\odot(E_\gamma)$ of the photons originating from the bremsstrahlung 
of the longitudinally polarized electrons on the radiator foil was computed using the Maximon-Olsen formula~\cite{maximon-olsen}.

\subsection{Particle Identification and Event Selection}

Particle identification for this analysis was based on time-of-flight. For each track of momentum $\vec{p}$, we compared the measured hadron flight time from 
the event vertex to the TOF system, $TOF_m$, to the expected time, $TOF_h$, for a pion, kaon, or proton of identical momentum. Cuts were placed on the difference 
between the measured and expected times, $\Delta TOF = TOF_m - TOF_h$. The particle species assumption that minimized $\Delta TOF$ was assigned as the particle type.
The CLAS TOF system timing resolution provided high efficiency for identification of $K^+$ and $p$ with minimal particle misidentification probability given the 
hadron momenta in this dataset. See Ref.~\cite{g12-note} for details.

The standard g12 analysis fiducial cuts and bad detector element knockout cuts were applied for this analysis~\cite{g12-note}. The effect of applying either tighter 
or looser fiducial cuts was studied as part of our systematic uncertainty analysis (see Section~\ref{sec-sys}). Additionally, the standard g12 hadron energy loss 
and momentum corrections were applied~\cite{g12-note}. The combined effect of these corrections is quite small. There is a slight narrowing of the $\Lambda$ peak 
in the $MM(K^+)$ distribution from $\sigma=11.7$~MeV to 10.8~MeV, along with a slight shift in the mean from 1.114~GeV to 1.116~GeV.

In this analysis we investigated two final state hadron topologies 

\begin{itemize}
\item Two-track events with a detected $K^+$ and $p$ 
\item Three-track events with a detected $K^+$, $p$, and $\pi^-$.
\end{itemize}

An initial data skim was used that required a detected $K^+$ in coincidence with at least one proton along with events that had a detected $K^+$ in coincidence 
with a $p$ and $\pi^-$. Events with an additional high momentum ($>2$~GeV) $\pi^+$ or a high momentum ($>3$~GeV) proton were also included in the selected event
sample in the skim since, at high momentum, there was insufficient resolution to separate them from kaons. 

\begin{figure}[htbp]
\centering
\includegraphics[width=0.85\columnwidth]{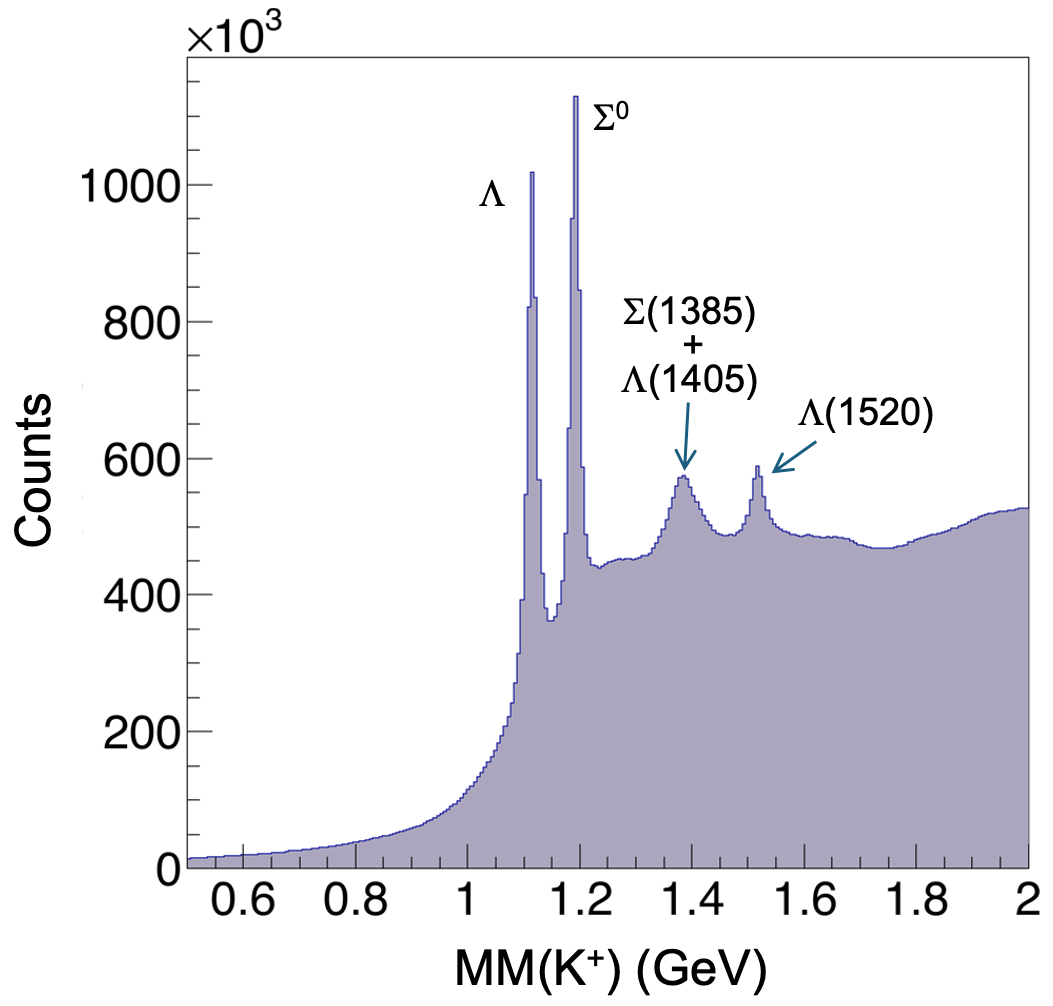}
\caption{Missing mass off the $K^+$ for three-track events that have one $K^+$, one $p$, and one $\pi^-$ before applying any cuts except those in the skim. Several
well-known hyperon states are labeled.} 
\label{fig-mm1}
\end{figure}

Exclusive $K^+ \Lambda$ events were identified by selecting the observed $\Lambda$ signal in the missing mass off the kaon, given by
\begin{equation}
\label{eq-mmK}
MM(K^+)=\sqrt{\left(P_\gamma+P_{tgt}-P_K\right)^2},
\end{equation}

\noindent
where $P_\gamma$, $P_{tgt}$, and $P_K$ are the four-momenta of the incident photon, target proton, and kaon, respectively. Figure~\ref{fig-mm1} shows 
the missing mass off the kaon for all events in the initial skim prior to any cuts.

Clearly, at this stage there is a large background under the $\Lambda$ peak. The remainder of this section details the cuts used to reduce the background and identify 
the final sample of exclusive $\Lambda$ events.

\subsection{Vertex Timing Cut}
\label{sec-vt}

The time difference between the start counter vertex time, $t_{SC}$, and the RF-corrected tagged electron vertex time, $t_{vtx}(TAG_{RF})$, was used to remove out-of-time
accidental events. $t_{vtx}(TAG_{RF})$ measures the arrival time of the photon at the event vertex and includes a correction for the propagation time, $t_{prop}$, 
from the target center to the event vertex
\begin{equation}
\label{eq-tTAG}
t_{vtx}(TAG_{RF})=t_{TAG,RF}+t_{prop},
\end{equation}
\noindent
where $t_{TAG,RF}$ is the RF-corrected time for the photon at the center of the target. The start-counter time also included a correction for the propagation 
of the particles from the event vertex to the start counter. The accidentals, which arose from particles produced in earlier or later beam bunches than the tagged 
electron, were removed with a $\pm 1.0$~ns cut placed on this difference (note: the beam bunches were separated by 2.004~ns).

\subsection{Vertex Position Cut}
\label{sec-vxyz}

\begin{figure}[htb!]
\centering
\includegraphics[width=0.97\columnwidth]{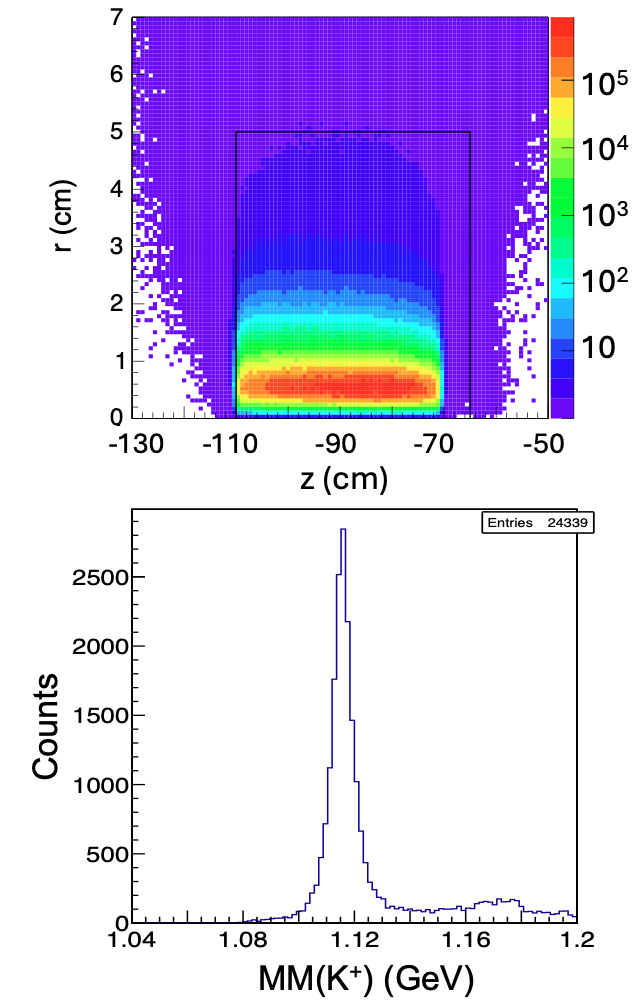}
\caption{Top: Hadron vertex position distribution in the $r$ vs. $z$ space for two-track events that pass all $K^+\Lambda$ cuts. The rectangular box shows the 
region kept in this analysis. Bottom: Missing mass off the $K^+$ for events that pass all cuts but with $-70\le z\le -65$~cm.} 
\label{fig:vertexDist}
\end{figure}

The cylindrical target for the g12 experiment had a radius of 2~cm and a length of 40~cm, and was positioned upstream of the nominal CLAS center from -110~cm 
to -70~cm. The best estimate of where an event occurred was the distance of closest approach to the beamline of the reconstructed tracks of the final-state 
particles. The top panel of Fig.~\ref{fig:vertexDist} shows the event-vertex radial position ($r=\sqrt{x^2+y^2}$) vs. the event vertex position along the 
beamline ($z$) for all events that satisfied our subsequent analysis cuts. We used the following cuts: $-110\le z\le -65$~cm and $r\le 5$~cm. The cuts extend 
beyond the outer radius and the downstream end of the target to include some $\Lambda$s that decay beyond the target. The bottom panel of Fig.~\ref{fig:vertexDist} 
shows that there were 25k events in the 5~cm range downstream of the target ($-70\le z\le -65$~cm). Compared to the roughly 10M events that end up in the final 
sample, the events we lose beyond the selected range are insignificant.

\subsection{Multiple Photon Cut}
\label{sec-mpc}

After applying the above timing cuts, $\sim$10\% of the events had more than one tagged photon. In the case of the multiple-photon events, rather than try to 
determine which photon was the correct photon, we simply removed all such events. This is a reasonable thing to do for asymmetry measurements since such 
a cut does not affect the extracted polarization. Furthermore, the loss of statistics is not significant.

\subsection{Kinematic Fitting}
\label{sec-kinfit}

\begin{figure}[htbp]
\centering
\includegraphics[width=0.8\columnwidth]{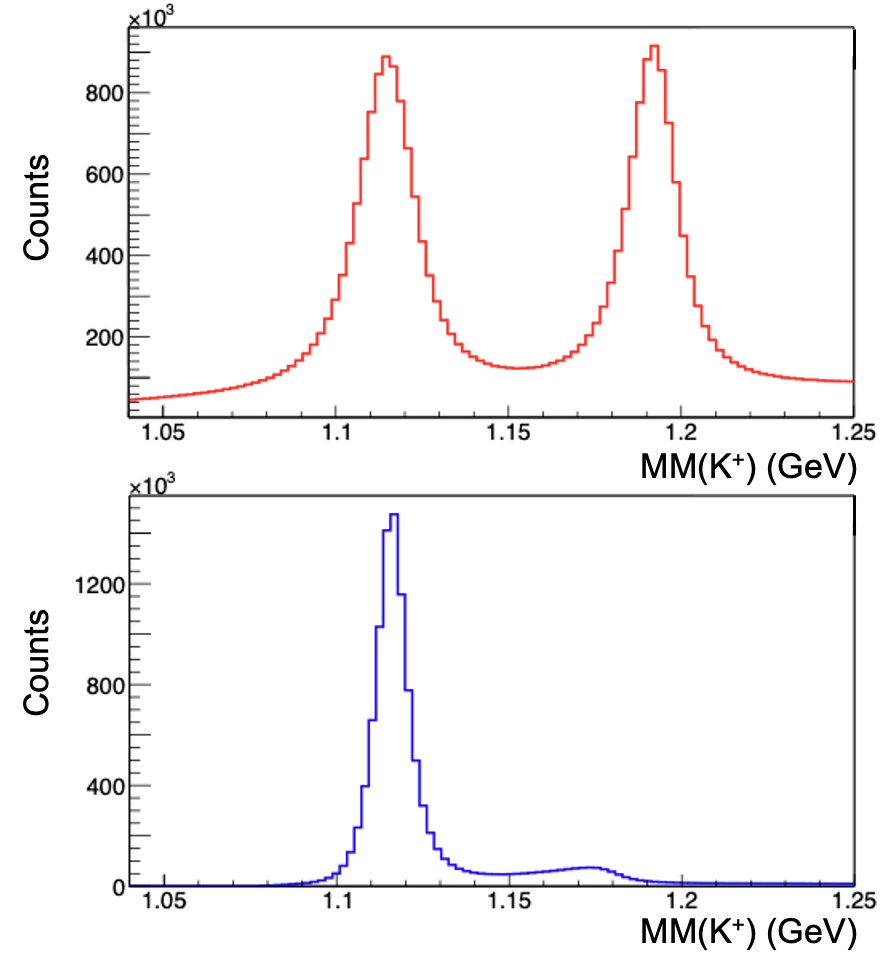}
\caption{Missing mass off the $K^+$ for the two-track events before (top) and after (bottom) kinematic fitting with all prior cuts and corrections.} 
\label{fig:mm2-track}
\end{figure}

The missing mass distribution for two-track events with all of the above cuts and corrections is shown in Fig.~\ref{fig:mm2-track} (top). Clear $\Lambda$ and $\Sigma^0$ 
peaks are seen along with a background from particle misidentification. We used kinematic fitting to reduce the background and select only the $K^+\Lambda$ events. The 
kinematic fitting process used the measured energy and momenta of the charged-particle tracks along with energy and momentum conservation constraints to improve the 
accuracy of the reconstructed track parameters. This analysis used the same procedure as previous CLAS analyses and is discussed elsewhere (see {\it e.g.} 
Ref.~\cite{KellerKinfit}).

In our analysis, we consider two final state hypotheses. One is the case of $\gamma p\to K^+ p (\pi^-)$, where the $\pi^-$ is missing so that the $\pi^-$ mass 
is the constraint. This is known as a single constraint (1-C) fit. The other hypothesis is $\gamma p \to K^+p\pi^-$, which has nothing missing and requires the
constraints of momentum and energy conservation. This is known as a four constraint (4-C) fit. 

\begin{figure}[htbp]
\centering
\includegraphics[width=0.81\columnwidth]{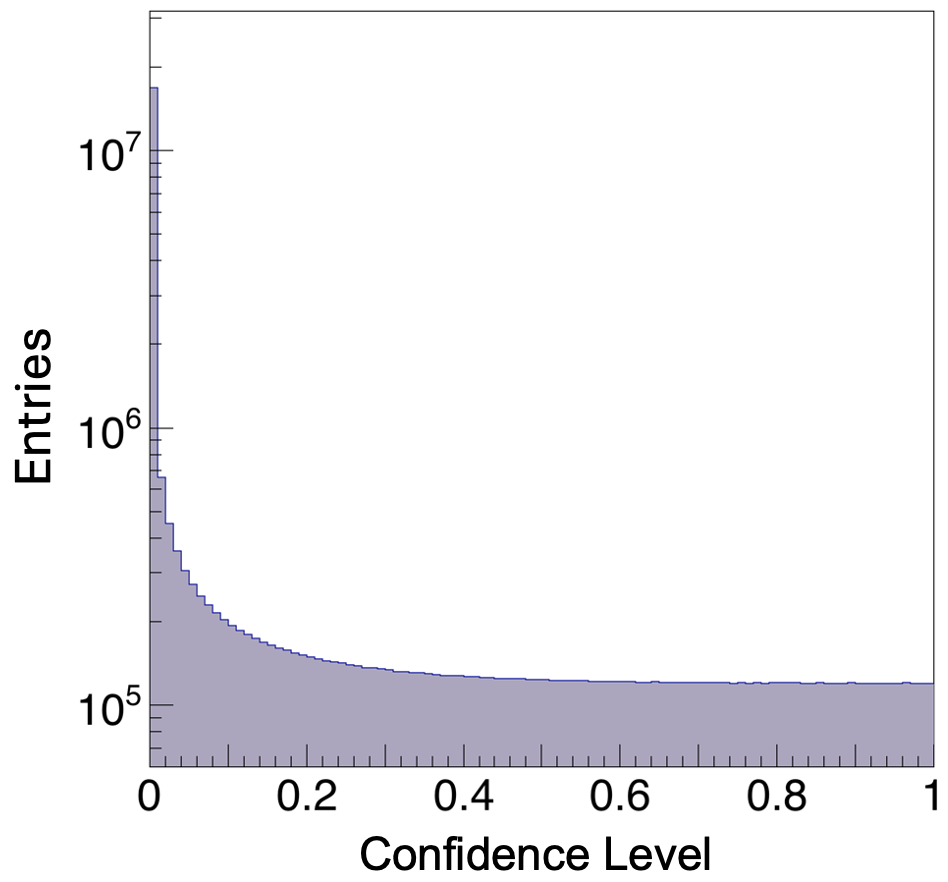}
\caption{Kinematic fitting confidence level distribution for the two-track events.}
\label{fig:CL}
\end{figure}

The key quantity that comes out of the kinematic fitter is the confidence level ($CL$), which takes into account the known energy and timing resolutions. In short, 
any fit that requires a measured quantity to move by a large amount relative to its resolution has a small confidence level. Figure~\ref{fig:CL} shows the confidence
level distribution for two-track events. Events that do not fit the final-state hypothesis show up at small values of $CL$, whereas those that do fit the final-state 
hypothesis lie in the relatively flat region. For the two-track and three-track events, the ``good event'' cuts are $CL>5$\% and 1\%, respectively. The 5\% confidence 
level cut is largely effective for background removal from the two-track analysis, as shown in Fig.~\ref{fig:mm2-track} (bottom). The effects of the small amount of 
remaining background are discussed in Section~\ref{sec-Q}. We studied the effect of varying the $CL$ cut to assign an associated systematic uncertainty (see 
Section~\ref{sec-cutsys}).

\begin{figure}[t]
\centering
\includegraphics[width=0.85\columnwidth]{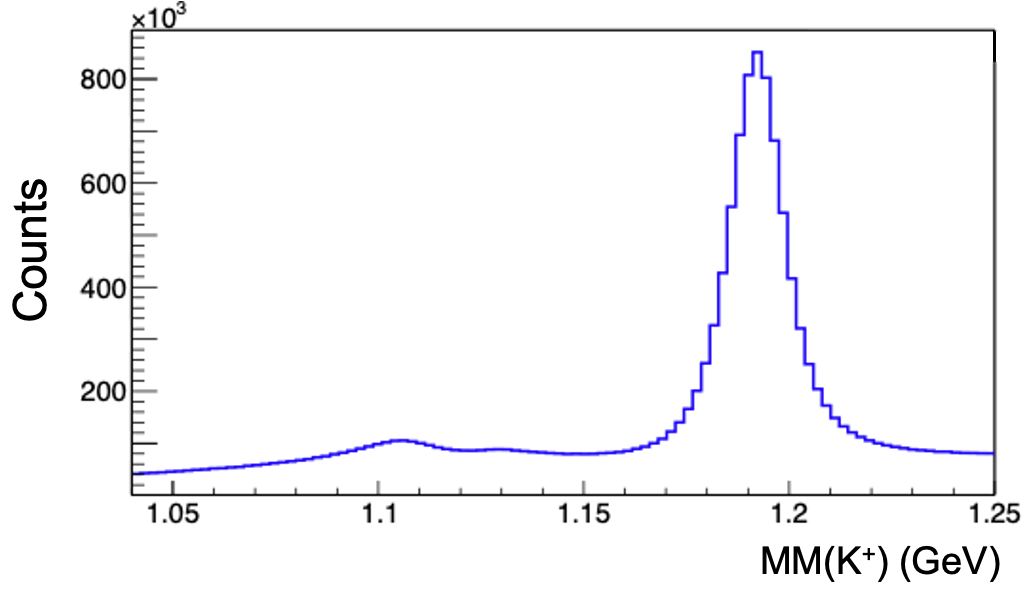}
\caption{Missing mass off the $K^+$ for two-track events rejected by the 5\% confidence level cut.}
\label{fig:background.CL} 
\end{figure}

Figure~\ref{fig:background.CL} shows the events rejected by the 5\% $CL$ cut for the two-track events. The $K^+\Sigma^0$ events are clearly rejected along with 
a relatively smooth background beneath the $\Lambda$ peak. A small number of $\Lambda$ events are also rejected (small hump at 1.1157~GeV), however their
rejection has no significant effect on the statistics of this analysis. 

\subsection{Background Rejection}
\label{sec-Q}

\begin{figure}[htb]
\centering
\includegraphics[width=0.95\columnwidth]{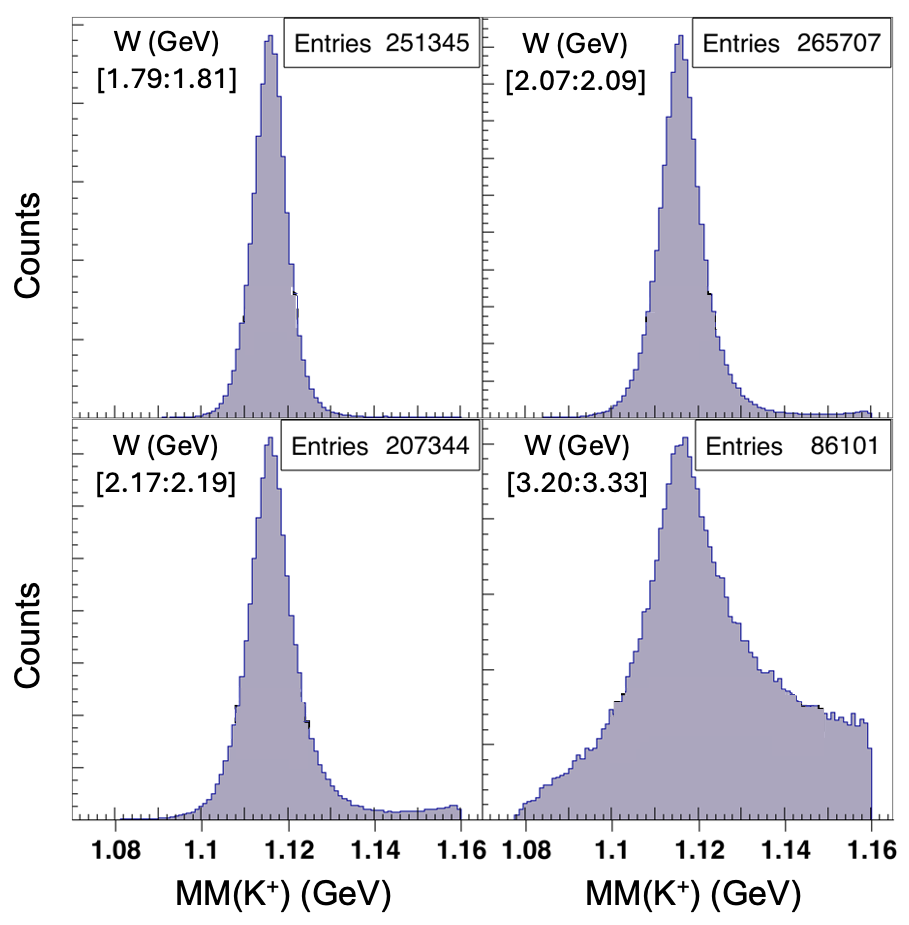}
\caption{Missing mass off the $K^+$ for two-track events after all cuts and kinematic fitting for $W$ bins as indicated.}
\label{fig:bgW}
\end{figure}

Figure~\ref{fig:mm2-track} shows that the kinematic fitting provides for a clean $\Lambda$ peak that has some background remaining. Figure~\ref{fig:bgW} shows 
that this background is relatively small at low $W$ but increases at higher $W$. This results from an increase in the particle misidentification probability at 
higher momenta due to the finite timing resolution of the TOF system.

In order to account for the background we have used the $Q$-factor method \cite{Williams:2008sh}, which has been used in previous CLAS analyses ({\it e.g.}
Refs.~\cite{mccracken2010,dey2010}). This method involves using nearest neighbors to assign each candidate $\Lambda$ event a quality factor 
(or $Q$-factor), which gives the probability that the event is a $\Lambda$. The $Q$-factors can then be used as event weights in the distributions used to extract 
the polarization observables (as mentioned in Section~\ref{sec-MLM}).

In this method, the data are assumed to be described by a set of coordinates. These coordinates are broken down into reference coordinates in which the 
signal and background distributions are assumed to be known or can be parameterized, and non-reference coordinates in which the signal and background
distributions are unknown. In our case, the reference coordinate is the missing mass ($MM$) off the kaon. We have assumed that the missing-mass distributions 
can be fit with a signal, $S(MM)$, consisting of a Breit-Wigner convoluted with a Gaussian ({\it i.e.} a Voigtian function), and a linear background function, 
$B(MM)$. Our non-reference coordinates are the kaon production angle ($\cos\theta_K^{c.m.}$), the invariant mass $W$, the decay proton angle in the $\Lambda$
rest frame ($\cos\theta_p^i$ for $i\in{x,y,z}$), and the azimuthal angle of the $p \pi^-$ decay plane relative to the $\gamma p \to K^+ \Lambda$ c.m. plane 
($\phi_p^\Lambda$).

\begin{figure}[htbp]
\centering
\includegraphics[width=1.0\columnwidth]{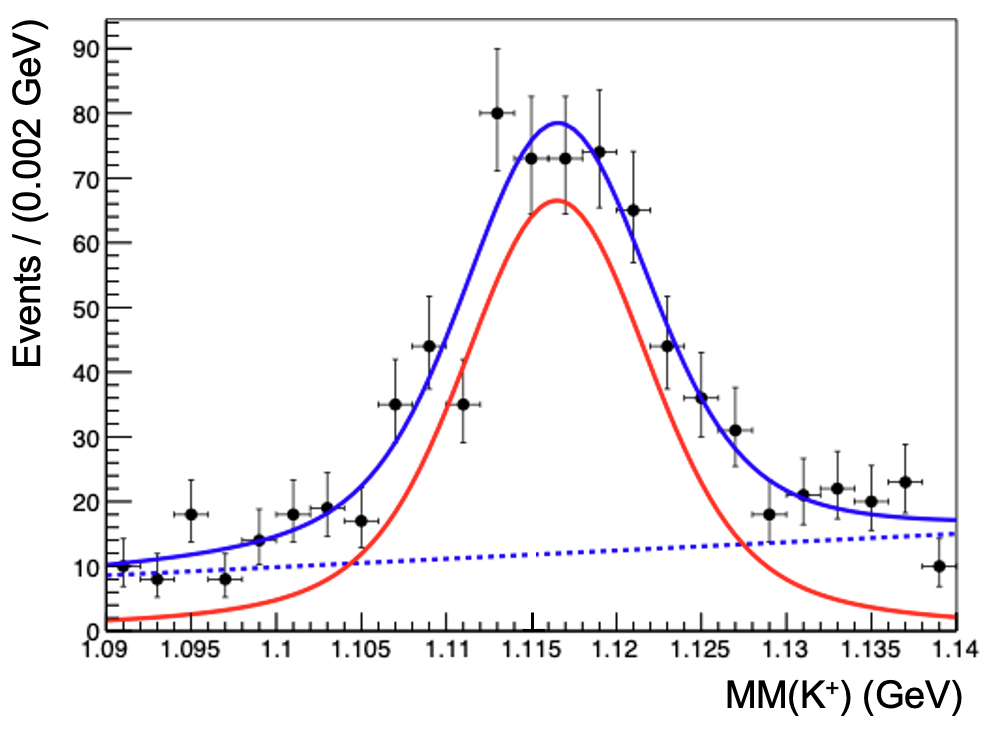}
\caption{Missing mass off the $K^+$ from 1000 nearest neighbor events for a randomly chosen. The red curve represents the signal fit, the blue dotted line represents 
the background fit, and the blue curve represents the total fit.}
\label{fig:qfactor.fits} 
\end{figure}

\begin{figure}[htbp]
\centering
\includegraphics[width=0.95\columnwidth]{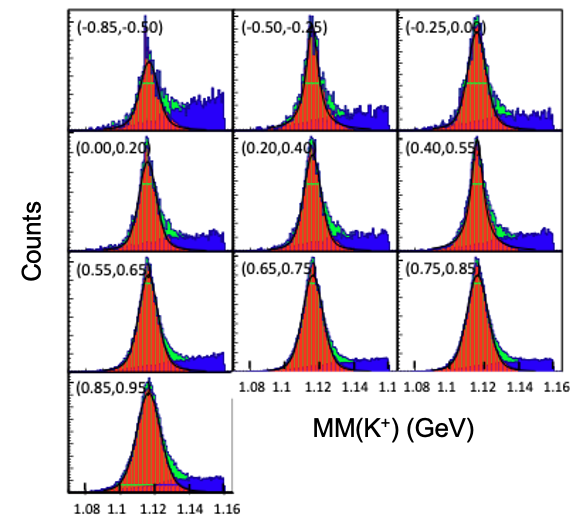}
\caption{Missing mass off the $K^+$ for a bin of $W \in [3.1, 3.2]$~GeV for $\cos\theta _K^{c.m.}$ bins as labeled with the $\Lambda$ signal ($Q$ values shown 
in orange) and the background ($1-Q$ values shown in blue) overlaid on the data (green).}
\label{fig:Qfitsample}
\end{figure}

For each event we then found the 1000 nearest neighbors in the phase space of the non-reference coordinates. These were identified by finding the distance from 
the target event given by
\begin{equation}
\label{eq-NNd}
d^2=\sum_{k=1}^4 \left( \frac{\xi_k^0-\xi_k}{r_k} \right)^{\!\!2},
\end{equation}
\noindent
where $\xi_k^0$ is the value of the non-reference coordinate of the target event, $\xi_k$ is the value of a neighboring event, and $r_k$ is the range of the 
non-reference coordinate (the bin width for $W$, $-1$ to $1$ for the cosine coordinates, and 0 to $2\pi$ for the azimuthal angle). Then for each target event a
missing-mass distribution for these 1000 nearest neighbors was fit with
\begin{equation}
\label{eq-sigBG}
f(MM)=f_s\cdot S(MM)+(1-f_s)\cdot B(MM),
\end{equation}
\noindent
where $f_s$ is the fitted signal fraction. The fitting procedure was done using an unbinned maximum likelihood method that is part of the {\it RooFit} package
\cite{Verkerke:2003ir}. Since the $Q$ value is the probability that an event is a signal event, it is given by
\begin{equation}
\label{eq-Qdef}
Q=\frac{f_s\cdot S(MM)}{f(MM)}.
\end{equation}
Figure~\ref{fig:qfactor.fits} shows an example of a fit of the 1000 nearest neighbors for a randomly chosen event. Figure~\ref{fig:Qfitsample} shows a set of
representative sample fits for one bin in $W$ overlaid with the $Q$-value (signal) and $1-Q$ (background) distributions.

\subsection{Kinematic Binning}
\label{sec-binning}

\begin{figure}[htbp]
\centering
\includegraphics[width=1.0\columnwidth]{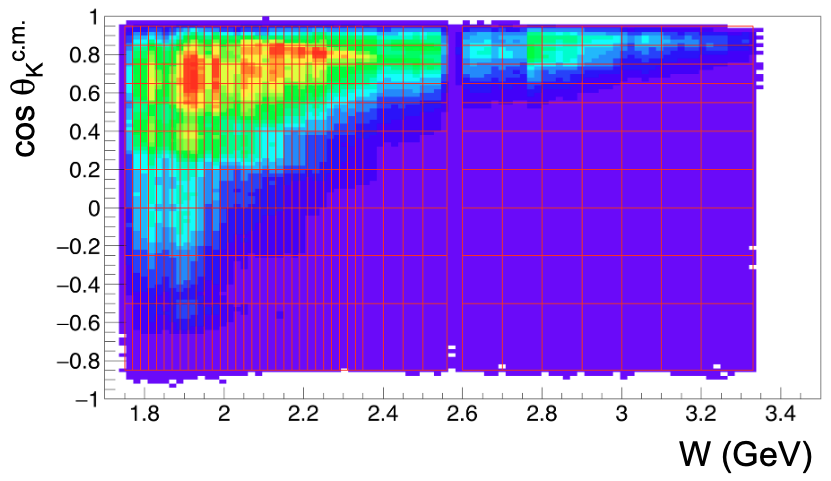}
\caption{Distribution of $\cos\theta_K^{c.m.}$ vs.~$W$ for two-track events with the bin limits overlaid. The depletion at $W \approx 2.6$~GeV is due to a bad tagger
detector element.}
\label{fig:bin}
\end{figure}

Our final results have been binned in $W$ and $\cos\theta_K^{c.m.}$ so that we can take advantage of the high statistics of the g12 data. Compared to the
previous CLAS results~\cite{bradford2007}, the $W$ binning is finer at low energies and wider at higher energies. We have broken the data into 11 bins in 
$\cos\theta_K^{c.m.}$, with larger bin sizes at back angles and smaller bin sizes at forward angles. Table~\ref{tab-bins} lists the binning ranges. Note that 
there is a gap in the data at $2.55 < W < 2.60$~GeV that results from a dead tagger paddle as seen in Fig.~\ref{fig:bin} that shows the $\cos \theta_K^{c.m.}$ 
vs. $W$ distribution for the two-track events with the bin-limits overlaid. Our results are reported using event-weighted bin centers.

\begin{table}[htbp]
\begin{center}
\begin{tabular}{|c|c|c|} \hline
$W$ Range (GeV)	& No. of Bins	& Width (MeV)\\ \hline
$[1.75, 2.35)$		& 30		& 20		 \\
$[2.35, 2.50)$		& 3			& 50		 \\
$[2.50, 2.56)$		& 1			& 60		 \\
$[2.60, 3.20)$		& 6			& 100		 \\
$[3.20, 3.33)$		& 1			& 130		 \\ \hline \hline
\multicolumn{3}{|c|}{$\cos\theta_K^{c.m.}$ Range}		\\ \hline
\multicolumn{3}{|c|}{$[-0.85, -0.65)$, $[-0.65, -0.45)$, $[-0.45, -0.25)$,}	\\
\multicolumn{3}{|c|}{$[-0.25, -0.05)$, $[-0.0, 0.15)$, $[0.15, 0.35)$,} \\
\multicolumn{3}{|c|}{$[0.35, 0.55)$, $[0.55, 0.65)$, $[0.65, 0.75)$,}				\\
\multicolumn{3}{|c|}{$[0.75, 0.85)$, $[0.85,0.95]$}					\\ \hline
\end{tabular}
\caption{Data binning for the two-track events in $W$ and $\cos \theta_K^{c.m.}$ for the $C_x$ and $C_z$ $\Lambda$ polarization analysis for the g12 data.}
\label{tab-bins}
\end{center}
\end{table}	

\subsection{Systematic Uncertainties}
\label{sec-sys}

This section reports the estimated systematic uncertainties for our nominal results pertaining to two-track events. All systematic uncertainties along with 
the totals are given in Table~\ref{tab:sys}. In the following subsections the different sources that contribute for this analysis are detailed.

\subsubsection{Cut-Related Uncertainties}
\label{sec-cutsys}

We have investigated several sources of point-to-point (uncorrelated) systematic uncertainty. These uncertainties associated with various event identification 
cuts were estimated by varying the cuts to produce an alternative value for the observables, ${\cal O}_{alt}$, and forming a weighted mean difference from the 
nominal values, ${\cal O}_{nom}$, according to

\begin{equation}
\label{eq-P2Psys}
\delta{\cal O}_{sys} = \sqrt{\frac{\displaystyle\sum _{i} \Big( \frac{\mathcal{O}_{nom}^{i} - \mathcal{O}_{alt}^{i}}
{\delta \mathcal{O}_{nom}^{i}}\Big)^{2}}{\displaystyle\sum_{i} \Big( \frac{1}{\delta \mathcal{O}_{nom}^{i}}\Big)^{2}}},
\end{equation}

\noindent
where $\delta {\cal O}_{nom}^i$ is the statistical uncertainty in ${\cal O}_{nom}^i$. The uncertainty obtained in this way was applied to all data points.

We note that this technique results in an overestimate of the systematic uncertainty since varying the cuts also changes the statistical uncertainty of the 
observable. However, the resulting systematic uncertainties are generally smaller--in most cases much smaller--than the statistical uncertainties.
The cut uncertainties are described below. 

\vskip 0.2cm
\noindent
{\bf Vertex timing cut}: We varied the vertex timing cut from $\pm 1.0$~ns to $\pm 0.9$~ns and $\pm 1.1$~ns. The estimated uncertainty is 0.003 for both 
$C_x$ and $C_z$.

\vskip 0.2cm
\noindent
{\bf Vertex $z$-position cut}: We produced a set of two alternate values for this study in which we changed the $z$-vertex cut by $\pm 10$\% from 
$-110\le z\le-65$~cm to $-112.25\le z \le 62.75$~cm and $-107.75\le z\le -67.25$~cm. The estimated uncertainties are 0.013 for $C_x$ and 0.012 for $C_z$.

\vskip 0.2cm
\noindent
{\bf Vertex $r$-position cut}: The alternative cuts for the radial vertex position were changed by $\pm 10$\% of the nominal cut from $r<5$~cm to 
$r<4.5$~cm and $r<5.5$~cm. The estimated uncertainties are 0.008 for $C_x$ and 0.007 for $C_z$.

\vskip 0.2cm
\noindent
{\bf Fiducial cuts}: We used both the standard g12 tight and loose fiducial cuts~\cite{g12-note} as alternative cuts for this study. The estimated 
uncertainties are 0.009 for $C_x$ and 0.008 for $C_z$.

\vskip 0.2cm
\noindent
{\bf Confidence level cuts}: We varied the confidence level cut from 5\% to 10\% as the alternative for this study. We only looked at an increased $CL$ 
cut since we wished to remain in the relatively flat portion of the $CL$ distribution. The resulting uncertainties are 0.017 for $C_x$ and 0.024 for $C_z$. 
In principle, increasing the $CL$ cut reduces the number of background events in our final sample so this is a partial measure of the effects of background. 
It is only a partial measure since there is still some background after applying a larger $CL$ cut so we also looked at the background effects through variations 
in the $Q$-value. 

\vskip 0.2cm
\noindent
{\bf $Q$-value and background uncertainty}: We estimated an additional background-related uncertainty by varying the $Q$ value for each event by the 
uncertainty in the $Q$ value. We then observed variations given by
\begin{equation}
\label{eq-Qdiff}
\Delta {\cal O}(Q_{\pm})={\cal O}_{nom}-{\cal O}_{\pm\delta Q},
\end{equation}
\noindent
where ${\cal O}_{\pm\delta Q}$ is the observable when we increase/decrease the $Q$ value by $\delta Q$. This variation lets in more background or cuts away 
more signal and thus accounts for background-related uncertainties. We then used both variations to calculate a difference used in Eq.(\ref{eq-P2Psys}). The 
resulting uncertainties are 0.013 for $C_x$ and 0.015 for $C_z$.

\vskip 0.2cm
\noindent
{\bf Acceptance effects -- two- vs. three-track events}: The statistical precision of the three-track results is not as good as for the two-track results. 
However, in the regions where the three-track results have relatively small statistical uncertainties--higher energies and forward angles--there the results
for both $C_x$ and $C_z$ are fully consistent. Since two and three-track events have very different acceptances, this agreement is strong evidence that 
the acceptance plays a minimal role in extracting $C_x$ and $C_z$. For $C_x$ and $C_z$ the mean of the difference distributions are nearly zero. As has been common 
for beam-polarization observables, we ignore acceptance effects in our systematic uncertainty budget and use the two-track/three-track comparison as a check on 
our assumptions.

\subsubsection{Scale-Type Uncertainties}
\label{sec-scalesys}

Scale-type uncertainties (or correlated uncertainties) affect all data in the same way and should not be added in quadrature with the point-to-point 
uncertainties. For this analysis the relevant sources are

\vskip 0.2cm
\noindent
{\bf Photon beam polarization}: This is determined as a relative uncertainty propagated from the electron beam polarization uncertainty. The electron 
beam polarization uncertainty is dominated by its systematic uncertainty and is $0.05 P_e$. Thus, $\delta C_i(P_\odot)=0.05|C_i|$ with $i = x,z$.

\vskip 0.2cm
\noindent
{\bf Weak decay asymmetry parameter}: The weak decay asymmetry parameter is $\alpha=0.747\pm 0.007$~\cite{pdg2024}, which has a relative uncertainty of 
$\delta\alpha/\alpha=0.009$, which is common to both $C_x$ and $C_z$.

\begin{table}[tbh]
\begin{center}
\begin{tabular}{|c|c|c|} \hline
	\multicolumn{3}{|c|}{Point-to-Point Uncertainties}\\ \hline
	Source			& $\delta C_x$		& $\delta C_z$ \\ \hline
	Timing cut			& 0.003			& 0.003		\\
	Vertex $z$ cut		& 0.013			& 0.012		\\
	Vertex $r$ cut		& 0.008			& 0.007		\\
	Fiducial cuts		& 0.009			& 0.008		\\
	Confidence level cut	& 0.017		& 0.024	\\
	$Q$-value			& 0.013			& 0.015		\\ \hline
	Total				& 0.028			& 0.033		\\ \hline\hline
	\multicolumn{3}{|c|}{Relative Scale-Type Uncertainties} \\ \hline
	$P_\odot$			& $0.05C_x$		& $0.05C_z$	\\
	$\alpha$			& $0.009C_x$		& $0.009C_z$	\\ \hline
	Total				& $0.051C_x$		& $0.051C_z$	\\ \hline
\end{tabular}
\caption{Systematic uncertainties for the $C_x$ and $C_z$ polarization analysis for the g12 data.}
\label{tab:sys}
\end{center}
\end{table}

\begin{figure*}[htbp]
\centering
\includegraphics[width=1.0\textwidth]{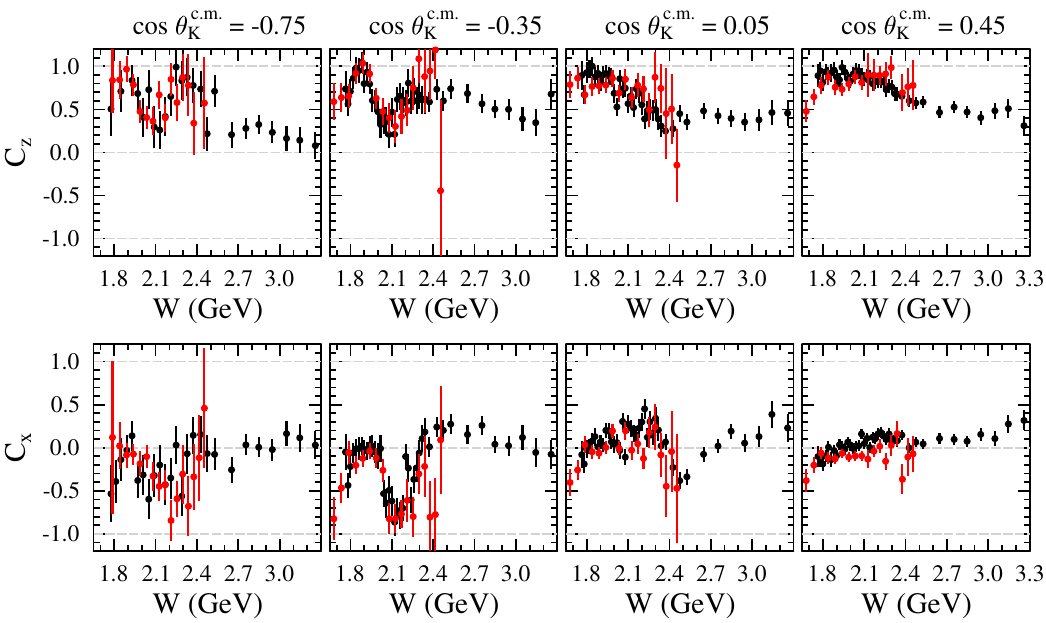}
\caption{$C_z$ (top) and $C_x$ (bottom) extracted from the earlier CLAS results~\cite{bradford2007} (red) compared to the present g12 results (black) from this 
work as a function of $W$ for representative $\cos \theta_K^{c.m.}$ bins as shown. The error bars include the quadrature sum of statistical and total (point-to-point
and scale-type) systematic uncertainties. The values are plotted for event-weighted $W$ values for angle bins as shown.}
\label{fig:g1cg12}
\end{figure*}

Figure~\ref{fig:g1cg12} compares our analysis results for hyperon polarization as a function of $W$ in representative $\cos \theta_K^{c.m.}$ bins with previously 
published results from CLAS~\cite{bradford2007}, scaled to account for the updated PDG value of $\alpha$. We see good overall agreement within uncertainties 
with the previous CLAS results over most of the kinematic range, but the new data provide for a significantly increased kinematic range and have much reduced 
statistical uncertainties. In addition the associated systematic uncertainties of the new analysis are in better control due to the use of kinematic fitting to 
reduce systematic uncertainties of the reconstructed hadron four-momenta, the use of the $Q$-factor method to improve the separation of the signal events from the 
background events, and the use of an unbinned log likelihood fit to extract the polarization observables.

\section{Results}
\label{sec:results}

\begin{figure*}[htbp]
\centering
\includegraphics[width=0.85\textwidth,trim={0 0.5cm 0 0},clip]{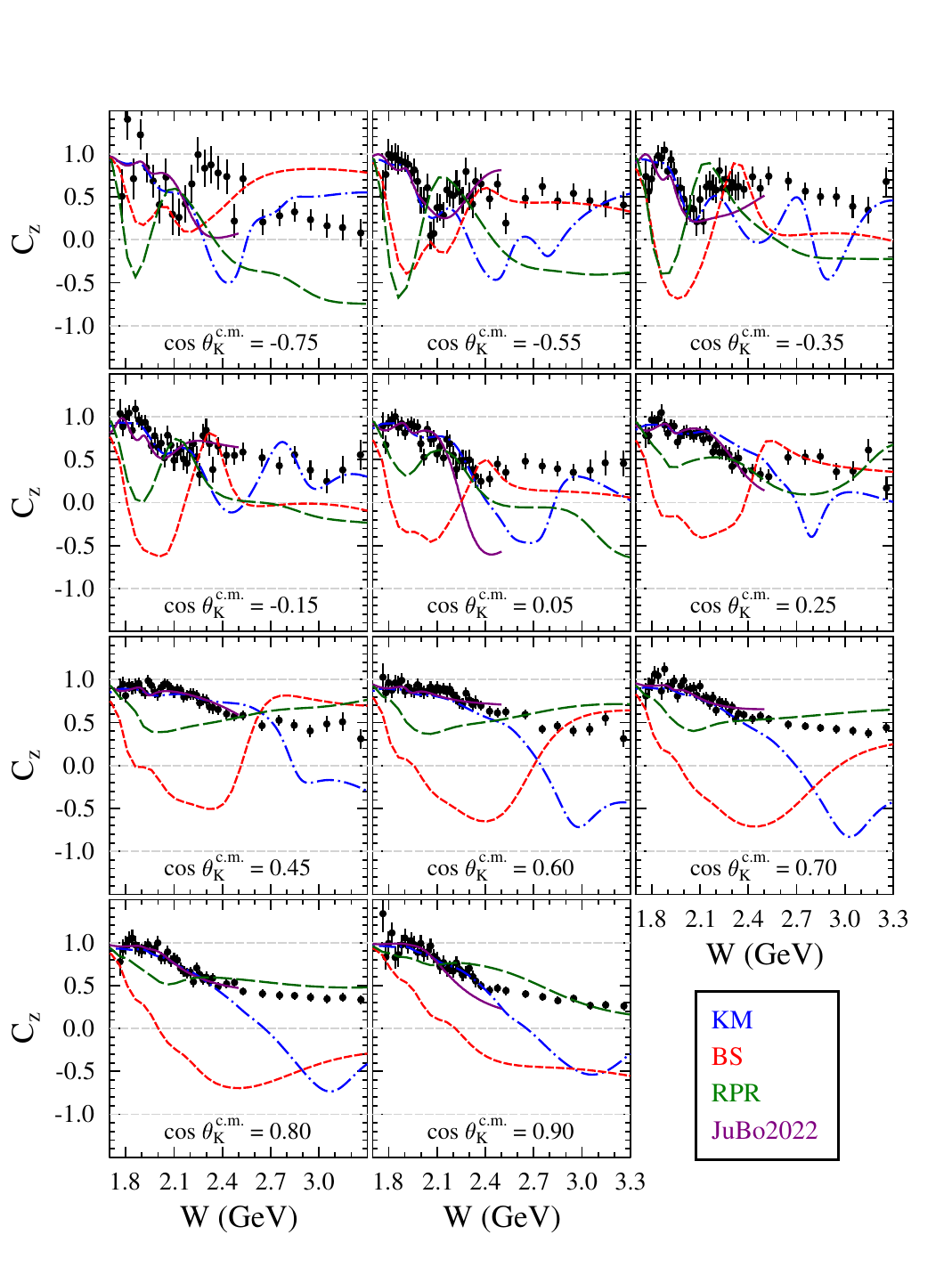}
\caption{Results for $C_z$ as a function of $W$ for different $\cos\theta_K^{c.m.}$ bins as indicated. The error bars include the quadrature sum of statistical and 
point-to-point systematic uncertainties. The values are plotted for event-weighted $W$ values for event-weighted angle bins. The curves are calculations from the 
isobar models of Kaon-Maid (KM)~\cite{kaon-maid1,kaon-maid2} (blue dot-dashed) and the Czech group (BS)~\cite{skoupil18} (red short dash), as well as the RPR 
model from the Czech group (RPR) \cite{rpr-bs} (green long dash). The coupled-channels solution from the JuBo2022 model~\cite{jubo2022} is shown by the purple solid 
curves.}
\label{fig:Czvsw}
\end{figure*}

\begin{figure*}[htbp]
\centering
\includegraphics[width=0.85\textwidth,trim={0 0.5cm 0 0},clip]{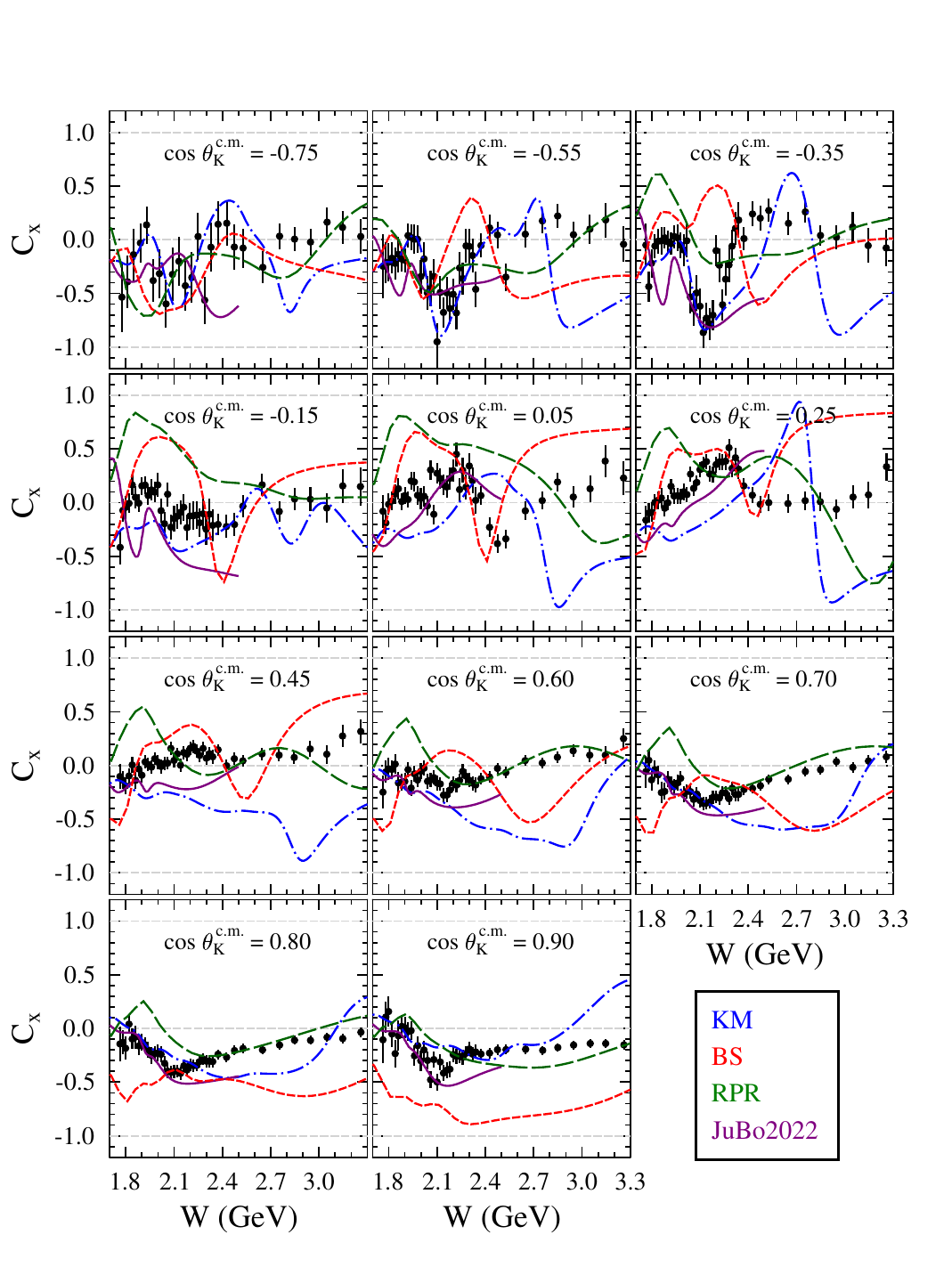}
\caption{Results for $C_x$  as a function of $W$ for different $\cos\theta_K^{c.m.}$ bins as indicated. The error bars include the quadrature sum of 
statistical and point-to-point systematic uncertainties. The values are plotted for event-weighted $W$ values for event-weighted angle bins. The model calculations 
are detailed in the caption of Fig.~\ref{fig:Czvsw}.}
\label{fig:Cxvsw}
\end{figure*}

\begin{figure*}[htbp]
\centering
\includegraphics[width=1.0\textwidth,trim={0 2cm 0 0},clip]{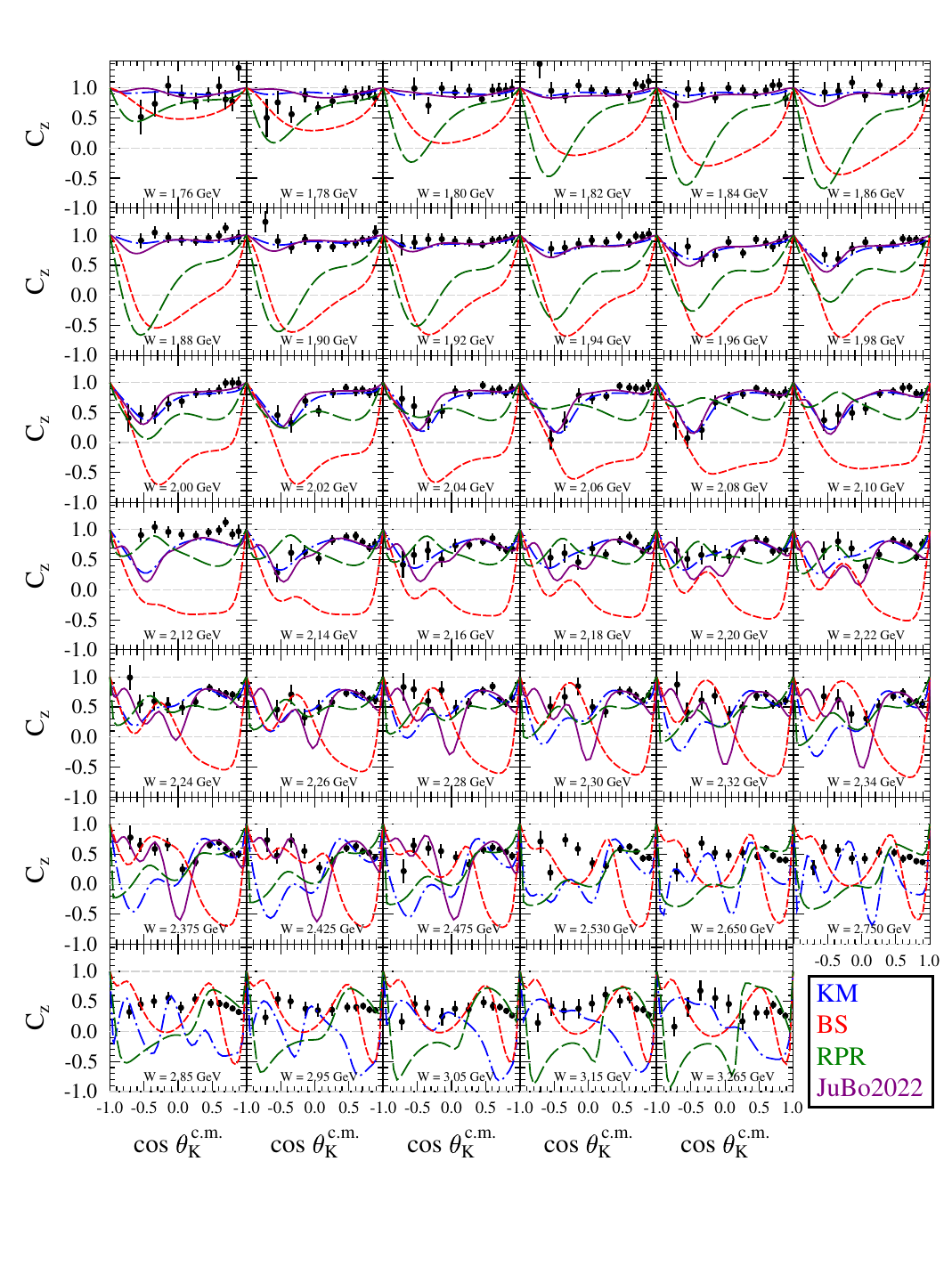}
\caption{Results for $C_z$  as a function of $\cos\theta_K^{c.m.}$ for different $W$ bins as indicated. The error bars include the quadrature sum of 
statistical and point-to-point systematic uncertainties. The values are plotted for event-weighted $\cos\theta_K^{c.m.}$ values for event-weighted angle bins. The model 
calculations are detailed in the caption of Fig.~\ref{fig:Czvsw}.}
\label{fig:Czvsc}
\end{figure*}

\begin{figure*}[htbp]
\centering
\includegraphics[width=1.0\textwidth,trim={0 2cm 0 0},clip]{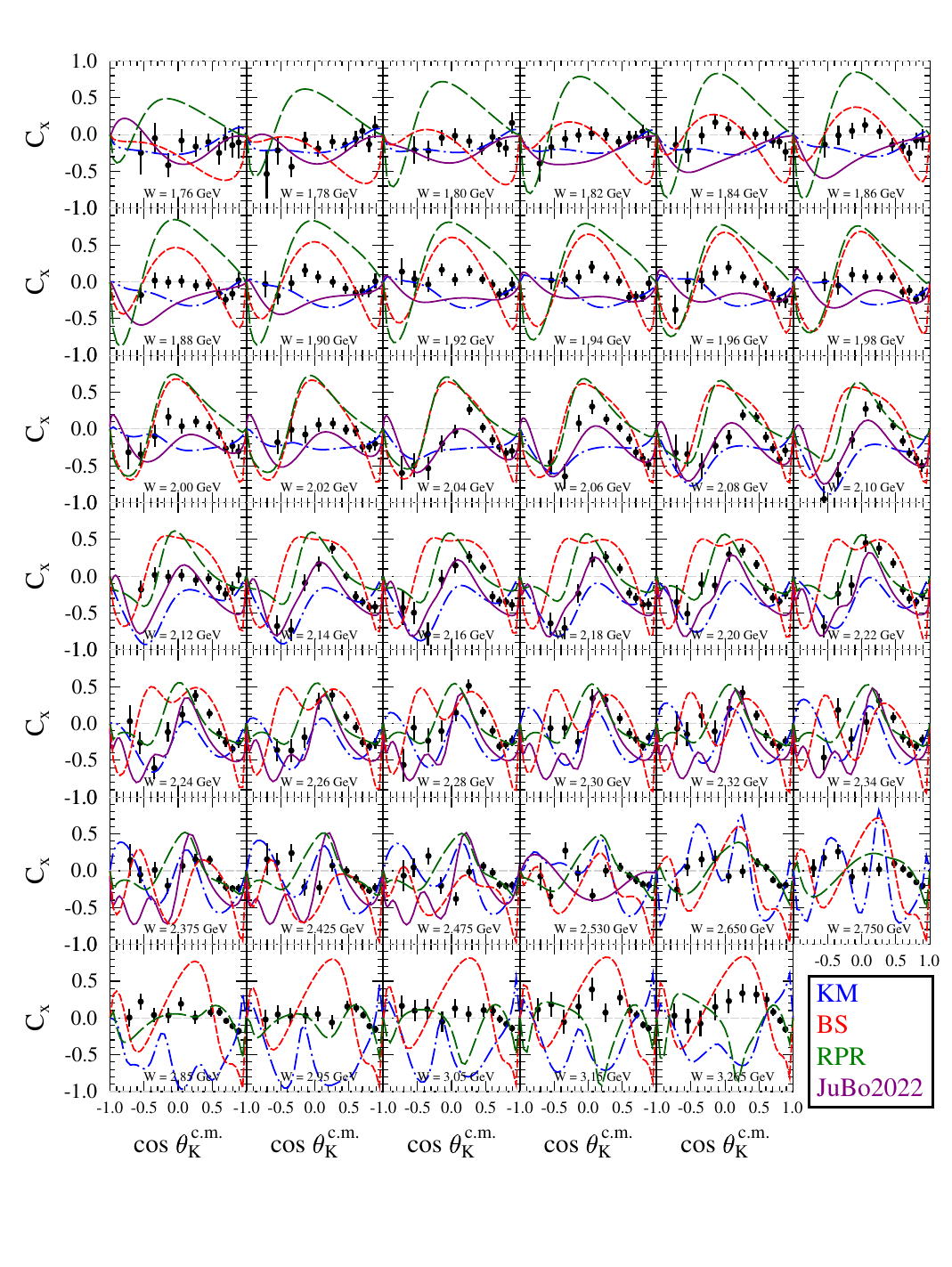}
\caption{Results for $C_x$  as a function of $\cos\theta_K^{c.m.}$ for different $W$ bins as indicated. The error bars include the quadrature sum of 
statistical and point-to-point systematic uncertainties. The values are plotted for event-weighted $\cos\theta_K^{c.m.}$ values for event-weighted angle bins. The model 
calculations are detailed in the caption of Fig.~\ref{fig:Czvsw}.}
\label{fig:Cxvsc}
\end{figure*}

In this section we present our results for $C_x$ and $C_z$. Our full set of results for $C_x$ and $C_z$ is shown as a function of $W$ for different bins of 
$\cos\theta_K^{c.m.}$ in Figs.~\ref{fig:Czvsw} and \ref{fig:Cxvsw}. They are also shown as a function of $\cos\theta_K^{c.m.}$ for different bins of $W$ in 
Figs.~\ref{fig:Czvsc} and \ref{fig:Cxvsc}. The results are included in the CLAS Physics Database~\cite{physicsdb}. 

The kinematic trends of the transferred $\Lambda$ polarization components $C_x$ and $C_z$ are already reasonably well understood within the resonance region based
on the older CLAS data from Ref.~\cite{bradford2007}. The photon polarization is largely transferred to the $\Lambda$ along the photon momentum direction ({\it i.e.} 
the $z$-axis) in the c.m.~frame. However, it has significant interference structure through the nucleon resonance region, particularly at backward $K^+$ c.m.~angles. 
The interference structures are significantly muted for forward $K^+$ c.m.~angles due to the dominance of $t$-channel exchange contributions. The polarization component 
$C_z$ is then seen to decrease in magnitude slowly and smoothly above $W \approx 2.4$~GeV for the full $K^+$ c.m. angle range. Where $C_z$ is large, the corresponding 
values of $C_x$ are close to zero as must be the case. Notably, just as is the case for $C_z$ at backward angles, $C_x$ too displays striking interference structures. 
However, at $W > 2.4$~GeV, $C_x$ is flat and very small in magnitude.

\vskip 0.2cm

There are several different models shown in this work to compare against the polarization observables. None of these models has been constrained by the present
$C_x$ and $C_z$ results. The main features of the models are discussed here to set the stage for their comparisons to the data.

\vskip 0.2cm
\noindent
\underline{Kaon-MAID} (blue dot-dashed curves in Figs.~\ref{fig:Czvsw} -- \ref{fig:Cxvsc}) - The Kaon-MAID (KM) model is a tree-level isobar model 
\cite{kaon-maid1,kaon-maid2} that includes Born terms, $K^*(892)$ and $K_1(1270)$ exchanges in the $t$-channel, and a limited set of spin 1/2, 3/2, and 5/2 
$s$-channel resonances. This core set includes the $N(1650)1/2^-$, $N(1710)1/2^+$, $N(1720)3/2^+$, and $N(1900)3/2^+$. These states were chosen as they 
have been reported to have non-zero decay widths into $K^+\Lambda$. The Born term vector meson exchange and the resonance couplings are based on fits to the 
available $\gamma p \to K^+Y$ and $\pi^-p \to K^0\Lambda$ data. The results included here were provided by its developer~\cite{mart-comm}.

The KM model reasonably matches the $C_x$ data for $W < 2.1$~GeV for backward $\cos \theta_K^{c.m.} < -0.35$ and for forward $\cos \theta_K^{c.m} > 0.60$ where
it has been constrained by the results of Ref.~\cite{bradford2007} but fares poorly otherwise. $C_z$ reasonably agrees with the data for the full angular range 
for $W$ up to $\approx 2.1$~GeV. The new data can clearly be seen to provide valuable information to improve the model constraints both within the nucleon resonance 
region and to higher $W$.

\vskip 0.2cm
\noindent
\underline{Byd\v{z}ovsk\'{y}-Skoupil} (red short-dash curves in Figs.~\ref{fig:Czvsw}--\ref{fig:Cxvsc}) - The Byd\v{z}ovsk\'{y}-Skoupil (BS) model~\cite{skoupil18} 
is another tree-level isobar model similar in design to the KM model. It is based on fits to some of the available $\gamma p \to K^+\Lambda$ photoproduction 
data (differential cross sections, recoil polarization, beam spin asymmetry). The full set of 3- and 4-star PDG $N^*$ and $\Delta^*$ resonances of spins up 
to 5/2 and $W$ up to 2~GeV are included. Like other isobar models, the BS model includes Born terms and exchanges in the $t$- and $u$-channels to account for 
the non-resonant backgrounds. In this work, the BS3 version of the model was used and the calculations were provided by its developers~\cite{czech-comm}.

The BS model does not compare well with the $C_x$ and $C_z$ data over the full kinematic phase space of the measurements. It reveals very strong interference
structures for $W < 2.4$~GeV that do not match even qualitatively to the data.
It will be important to include these new polarization observables into the model
constraints.

\vskip 0.2cm
\noindent
\underline{RPR} (green long dash curves in Figs.~\ref{fig:Czvsw} -- \ref{fig:Cxvsc}) - A hybrid Regge plus resonance model (RPR) was developed by the Czech group
\cite{rpr-bs}. It is based on an isobar model to describe the $s$-channel resonance contributions and a Regge model to constrain the non-resonant backgrounds. 
The Regge modeling of the backgrounds involves the exchange of $K$ and $K^*$ trajectories. The model includes nearly a dozen $N^*$ states in the mass range up 
to 2.6~GeV for states of spin up to 5/2. A novel feature of this RPR model compared to the older RPR model from the Ghent group~\cite{rpr} is in its scheme for 
gauge invariance restoration. This model is based on fits that include the previously available CLAS $\gamma p \to K^+\Lambda$ cross sections and recoil 
polarization data. The RPR model variant included in this work employs a pseudovector coupling in the $pK\Lambda$ coupling. The RPR calculations were provided 
by its developers~\cite{czech-comm}.

The RPR model does not compare well with the data over the majority of the kinematic range in $W$ or $\cos \theta_K^{c.m.}$. It reveals very strong interference
structures in $C_x$ and $C_z$ that do not track the data. However, the model does reasonably produce the trends of $C_x$ and $C_z$ for forward 
$\cos \theta_K^{c.m.} > 0.70$ for $W > 2.4$~GeV, which is not unexpected due to the dominance of $t$-channel contributions in this range of kinematics that are
well described by Reggeon exchanges. Of course, these new data that extend beyond the resonance region can serve as an important additional constraint to the 
non-resonant contributions that extend down into the resonance region.

\vskip 0.2cm

The comparisons of the predictions of the RPR, KM, and BS models to the data make clear that these new results for $C_x$ and $C_z$ should be included in the 
fitting database for these models to constrain the $s$-channel resonance set, the associated resonance parameters, and the coupling constants. Given the quality 
of these new data, they can be expected to be quite valuable in this regard.

\vskip 0.2cm
\noindent
\underline{J\"ulich-Bonn} (purple solid curves in Figs.~\ref{fig:Czvsw} -- \ref{fig:Cxvsc}) - The J\"ulich-Bonn (JB) coupled-channel model was initially developed 
to study hadronic channels formed through $\pi N$ interactions~\cite{jb-model}. It was extended to include most of the available CLAS $K^+\Lambda$ cross section 
and polarization data for photoproduction~\cite{jb-kl1}. In this approach two-body unitary and analyticity are respected. The JuBo2022 model for $K^+\Lambda$ 
included here is limited to $W_{max} = 2.5$~GeV~\cite{jubo2022}. The calculations were provided by one of its developers~\cite{ronchen-comm}.

Comparisons of JuBo2022 to the data show reasonable agreement over the full kinematic range for $W < 2.1$~GeV for $\cos \theta_K^{c.m.} > 0.6$ but the model fares 
poorly in detail elsewhere. However, the coupled-channels fits do at least match the sign of the polarization for the region of $\cos \theta_K^{c.m.} < 0.6$. A next 
step to improve the JuBo2022 model results would be to include these new CLAS data given the valuable new constraints that they supply.

\vskip 0.2cm

There are several inequalities that must be satisfied by the observables in pseudoscalar meson photoproduction. For a circularly polarized beam it must be that
\cite{ars07}

\begin{equation}
\label{rlam}
R_\Lambda^2 \equiv P^2 + C_x^2 + C_z^2 \le 1,
\end{equation}

\noindent
where $P$, $C_x$ and $C_z$ are the polarization observables defined in Eq.(\ref{eq-AllP}). Thus the magnitude of the three orthogonal polarization components may have 
any value up to unity. As detailed in Ref.~\cite{bradford2007}, Eq.(\ref{rlam}) does not require that the $\Lambda$ is produced fully polarized except at 
$\vert \cos \theta_K^{c.m.} \vert = 1$ where orbital angular momentum plays no role. In Ref.~\cite{bradford2007} the value of $R_\Lambda$ was shown to be consistent 
with unity to within the available data uncertainties over the full range of the kinematics explored in that work, $-0.85 < \cos \theta_K^{c.m.} < 0.95$ and $W$ up to 
2.3~GeV. It was concluded that $\Lambda$ hyperons produced with circularly polarized photons appear to be 100\% spin polarized. As this is not required by the kinematics 
of the process, there must be some as yet unknown dynamical explanation. 

\begin{figure*}[htbp]
\centering
\includegraphics[width=1.0\textwidth]{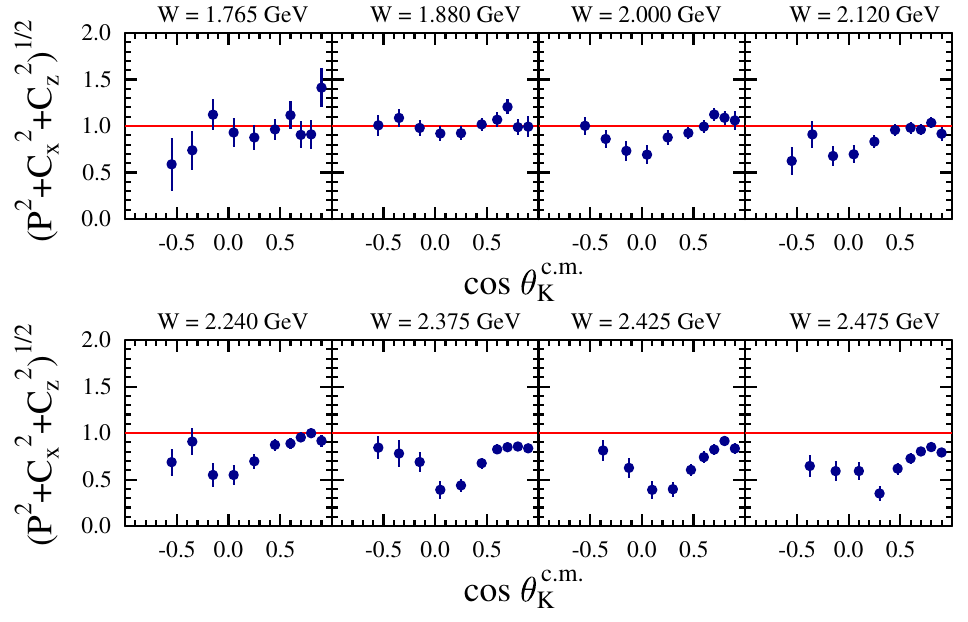}
\caption{Magnitude of the $\Lambda$ hyperon polarization vector $R_\Lambda = \sqrt{P^2 + C_x^2 + C_z^2}$ as a function of $\cos \theta_K^{c.m.}$ for selected bins in 
$W$ as labeled. The error bars shown are the quadrature sum of the statistical and systematic uncertainties of the measurements.}
\label{fig:polsum}
\end{figure*}

Using the measured $C_x$ and $C_z$ data from this work, we can further explore $R_\Lambda$ taking the measured $P$ values from the existing CLAS results of 
Ref.~\cite{mccracken2010}. These $C_x$, $C_z$, and $P$ data have much reduced uncertainties compared to what was used in Ref.~\cite{bradford2007} to compute
$R_\Lambda$. As the binning of Ref.~\cite{mccracken2010} is slightly different than employed in this work, a linear interpolation procedure was used to evolve the 
$P$ values to match the bins of $C_x$ and $C_z$ presented here. The results for $R_\Lambda$ for selected $W$ values up to 2.475~GeV are shown in Fig.~\ref{fig:polsum}. 
Here it is seen that $R_\Lambda$ is consistent with unity for $W$ up to $\sim$1.9~GeV and then begins to fall off in average value with a pronounced dip in 
$\cos \theta_K^{c.m.}$ in the range from 0 to 0.2. The dynamical origin of this striking energy dependence of the $\Lambda$ polarization calls for understanding.
This point was developed further in Ref.~\cite{schumacher2008} developing a simple picture in terms of quark dynamics to explain the data trends.

\section{Conclusions}
\label{sec:conclusions}

In summary, we have presented results from the CLAS detector for the beam-recoil polarization observables $C_x$ and $C_z$ for $\Lambda$ photoproduction from the 
proton in the exclusive $K^+\Lambda$ reaction in the energy range from threshold to above the nucleon resonance region up to $W=3.33$~GeV spanning the full $K^+$
c.m.~angular range. The current work is an extension of previously published CLAS photoproduction data that went up to $W=2.5$~GeV with significantly improved 
statistical precision in the region of overlap. It is notable that $C_z$ for the $\Lambda$ is large and positive, indeed near $+1.0$, over the range of $W$ in the 
nucleon resonance region up to $W \approx 2$~GeV. Including the existing $\Lambda$ recoil polarization data from CLAS, the data show that the $\Lambda$ hyperon is 
produced fully polarized in this $W$ range for a circularly polarized beam and then its polarization magnitude begins to fall off for increasing $W$. 

No existing models describe the data well over the full kinematic range of the new data, so we expect that reconsideration of these models in view of these new 
results may lead to new insights into the dynamics of strange quark photoproduction and the contributing $N^*$ states that couple to $K^+\Lambda$. The extended 
coverage to higher $W$ will enable improved understanding of the non-resonant backgrounds that extend from above the nucleon resonance region down into the domain 
where the $N^*$ resonances populate. It remains unclear whether the discrepancy between the new measurements and the existing models arises from the omission of 
higher-mass resonances or whether it could be resolved through improved constraints on the parameters of the resonant states and background terms already included 
in the models. An important next step is to include these new CLAS data into the model fits. The new data with their improved kinematic reach and significantly 
improved statistical precision compared to the existing CLAS data~\cite{bradford2007} should afford an important opportunity to significantly improve the parameter 
constraints and to further understand the shortcomings of the available models. In the longer term, for a more complete understanding of $K^+\Lambda$ photoproduction,
additional observables, including differential cross sections, over the new $W$ range will be crucial to better constrain the model parameters.

\begin{acknowledgments}

We thank Petr Byd\v{z}ovsk\'{y}, Terry Mart, Deborah R\"onchen, and Dalibor Skoupil for their efforts in preparing the model calculations for this paper. The authors 
would like to acknowledge the outstanding efforts the staff of the Accelerator and the Physics Divisions at Jefferson Lab in making this experiment possible. This work 
was supported in part by the U.S. Department of Energy, the National Science Foundation (NSF), and
the Italian Istituto Nazionale di Fisica Nucleare (INFN), 
the French Centre National de la Recherche Scientifique (CNRS), 
the French Commissariat pour l'Energie Atomique, 
the UK Science and Technology Facilities Council (STFC).
the National Research Foundation (NRF) of Korea, 
the Chilean Agencia Nacional de Investigacion y Desarrollo (ANID),
and the Skobeltsyn Nuclear Physics Institute and Physics Department at the Lomonosov Moscow State University. This work was supported in part by the the U.S. Department 
of Energy, Office of Science, Office of Nuclear Physics under contract DE-AC05-06OR23177.

\vskip 0.2cm
\noindent
$^\dag$Corresponding authors: {\small carman@jlab.org, baraue@fiu.edu}

\end{acknowledgments}

\end{document}